\documentclass[%
 preprint,
 superscriptaddress,
 amsmath,amssymb,
 aps,
 prb,
 floatfix,
]{revtex4-2}

\usepackage{graphicx}
\usepackage{dcolumn}
\usepackage{bm}
\usepackage{hyperref}

\setcitestyle{super}

\usepackage{xr}

\usepackage[usenames,dvipsnames]{color}
\usepackage{colortbl}
\usepackage[english]{babel}
\newcommand{\REV}[1]{#1} 
\newcommand{\DEL}[1]{}

\begin{document}

\title{Attosecond current control and timing in a scanning tunnelling microscope}

\author{Daniel Davidovich}
\altaffiliation{These authors contributed equally to this work.}
\affiliation{Department of Physics, Technion---Israel Institute of Technology, Haifa 32000, Israel}
\affiliation{Solid State Institute, Technion---Israel Institute of Technology, Haifa 32000, Israel}
\affiliation{The Helen Diller Quantum Center, Technion---Israel Institute of Technology, Haifa 32000, Israel}

\author{Boyang Ma}
\altaffiliation{These authors contributed equally to this work.}
\affiliation{Department of Physics, Technion---Israel Institute of Technology, Haifa 32000, Israel}
\affiliation{Solid State Institute, Technion---Israel Institute of Technology, Haifa 32000, Israel}
\affiliation{The Helen Diller Quantum Center, Technion---Israel Institute of Technology, Haifa 32000, Israel}

\author{Adi Goldner}
\altaffiliation{These authors contributed equally to this work.}
\affiliation{Department of Physics, Technion---Israel Institute of Technology, Haifa 32000, Israel}
\affiliation{Solid State Institute, Technion---Israel Institute of Technology, Haifa 32000, Israel}
\affiliation{The Helen Diller Quantum Center, Technion---Israel Institute of Technology, Haifa 32000, Israel}

\author{Shimon Cohen}
\affiliation{Department of Physics, Technion---Israel Institute of Technology, Haifa 32000, Israel}
\affiliation{Solid State Institute, Technion---Israel Institute of Technology, Haifa 32000, Israel}
\affiliation{The Helen Diller Quantum Center, Technion---Israel Institute of Technology, Haifa 32000, Israel}

\author{Zhaopin Chen}
\affiliation{Department of Physics, Technion---Israel Institute of Technology, Haifa 32000, Israel}
\affiliation{Solid State Institute, Technion---Israel Institute of Technology, Haifa 32000, Israel}
\affiliation{The Helen Diller Quantum Center, Technion---Israel Institute of Technology, Haifa 32000, Israel}

\author{\REV{Andrei G. Borisov}}
 \affiliation{Institut des Sciences Mol\'eculaires d'Orsay (ISMO), UMR 8214, CNRS, Universit\'e Paris-Saclay, 91405 Orsay Cedex, France.}
      \affiliation{Donostia International Physics Center, Manuel de Lardizabal 4, 20018 Donostia, Spain}

\author{Michael Kr\"uger}
\altaffiliation{Corresponding author: krueger@technion.ac.il}
\affiliation{Department of Physics, Technion---Israel Institute of Technology, Haifa 32000, Israel}
\affiliation{Solid State Institute, Technion---Israel Institute of Technology, Haifa 32000, Israel}
\affiliation{The Helen Diller Quantum Center, Technion---Israel Institute of Technology, Haifa 32000, Israel}

\date{\today}



\maketitle



\textbf{Quantum tunnelling of electrons can be confined to the sub-cycle time scale of strong light fields, contributing decisively to the extreme time resolution of attosecond science. Because tunnelling also enables atomic-scale spatial resolution in scanning tunnelling microscopy (STM), integrating STM with light pulses has long been a key objective in ultrafast microscopy, spanning the picosecond and femtosecond domains, with first signatures of attosecond dynamics. However, while sub-cycle dynamics on the attosecond time scale are routinely controlled and determined with high precision, controlling the direction of attosecond currents and determining their duration have remained elusive in STM. Here, we induce STM tunnelling currents using two-colour laser pulses and dynamically control their direction, relying solely on the sub-cycle waveform of the pulses. 
\REV{Projecting our measurement data onto one-electron and many-body theory descriptions reveals a three-step transport process in the non-adiabatic tunnelling regime as the physical mechanism, with a theory-derived current burst duration of 860\,as.}
Despite working under ambient conditions but free of thermal artifacts, we achieve sub-angström topographic sensitivity and a lateral spatial resolution of 2\,nm. This unprecedented capability to directionally control attosecond bursts will enable triggering and imaging ultrafast charge dynamics at the spatio-temporal microscopy frontier of lightwave electronics.}
\\
Attosecond science is based on the control of electron motion on the atomic time scale by strong electric fields~\cite{Corkum2007}. While it has enabled the observation of a wide range of attosecond phenomena, achieving a spatial resolution on the order of the atomic length scale in the angstr\"om-nanometer regime is challenging. The main pathway to attosecond microscopy is the use of attosecond-scale electron pulses instead of light due to their small de Broglie wavelength, typically much less than 1\,nm. In addition, electrons must be spatially confined, either as a recolliding electron wavepacket originating from an atom, molecule, or nanostructure~\cite{Niikura2002,Pullen2015,Kruger2011,Dienstbier2023,Kim2023}, as a narrow beam in an electron microscope~\cite{Priebe2017,Nabben2023}, or inside an atomic-scale tunnelling microscopy (STM) junction~\cite{Garg2020,Garg2022,Muller2023}. The strength of the latter approach is that STM in its static implementation using a bias field provides atomic resolution and precise energy resolution out of the box for a wide range of systems, such as molecules~\cite{Repp2005}. STM tunnelling currents driven by single-cycle THz transients~\cite{Cocker2013}, which can be regarded as a slowly varying static bias field, have enabled real-time observations of atomic-scale molecular vibrations on the sub-picosecond time scale~\cite{Cocker2016} and exciton formation in molecules~\cite{Kimura2025}, for instance.

Tuning the carrier-envelope phase (CEP) can shape an ultrashort THz field waveform to break its symmetry and obtain a field that is much stronger in one direction compared to the other. Maximum tunnelling is then achieved by the field in that direction, enabling control of the dominant direction of the current. This allows, for example, the controlled extraction of an electron from the sample~\cite{Yoshioka2016}, an important prerequisite for pump-probe measurements~\cite{Cocker2016}. Moving to infrared wavelengths and the femtosecond and attosecond time scales, CEP modulation is a well-known approach to control the sub-cycle electron motion in atoms~\cite{Baltuska2003,Paulus2004,Uiberacker2007}, metallic nanotips~\cite{Kruger2011, Piglosiewicz2014, Kim2023}, nanodevices~\cite{Bionta2021} and vacuum nanogaps~\cite{Rybka2016, Ludwig2020, Luo2023}\REV{, enabling} attosecond time resolution in these systems. It may seem straightforward to induce attosecond tunnelling currents and control their direction in STM using CEP-stable infrared laser pulses, building on many experiments exploring ultrafast STM in the femtosecond regime (see, e.g.,~\cite{Gerstner2000,Takeuchi2004,Lee2010,Dolocan2011,Schroder2020,Heimerl2020,Muller2023}). However, pioneering ultrafast STM experiments with infrared lasers have shown that thermal artifacts are often present when conventional lock-in approaches are used to extract low-level laser-induced currents~\cite{Gerstner2000,Gerstner2000a}. First signatures of CEP-controlled tunnelling current modulation in STM using near-infrared \REV{two-cycle~\cite{Garg2020} and single-cycle pulses~\cite{Rossetti2025} were reported recently}. Still, \REV{a direct demonstration of sub-cycle control of the direction of STM currents, together with theory-constrained timing of their dynamics, has remained elusive}. This advance is necessary for attosecond imaging of ultrafast electron phenomena.

In this work, we demonstrate robust attosecond directional control of ultrafast tunnelling currents in an STM junction. We employ two-colour pulses, a powerful alternative to CEP-stable pulses, and achieve ultrafast measurements without thermal artifacts. To this end, we superimpose an infrared laser pulse with its second harmonic in the same polarization plane and change their relative time delay. Two-colour pulses lead to field waveform\REV{s} with controlled symmetry breaking and enable\DEL{s} sub-cycle control, as demonstrated in a wide range of attosecond phenomena~\cite{Schumacher1994,Dudovich2006,Vampa2015a,Forster2016,Dienstbier2023}. Using the synthesized field waveform of the two-colour pulse, we are able to seamlessly switch between tip-to-sample tunnelling and vice versa\REV{, as shown unambiguously and directly by measurements with a temporarily frozen tip}. From an excellent agreement of experiment and theory we are able to identify the underlying physical mechanism as non-adiabatic tunnelling, with \REV{theory-derived} electron burst durations of $\sim$\,$860$\,as. \REV{Applying an artifact-free lock-in approach, our two-colour-driven microscope enables a lateral spatial resolution of 2\,nm despite operation under ambient conditions.} 

\section*{Results}

\subsection*{Ultrafast two-colour scanning tunnelling microscopy}

\begin{figure}[htb!]
    \centering
    \includegraphics[width=1\linewidth]{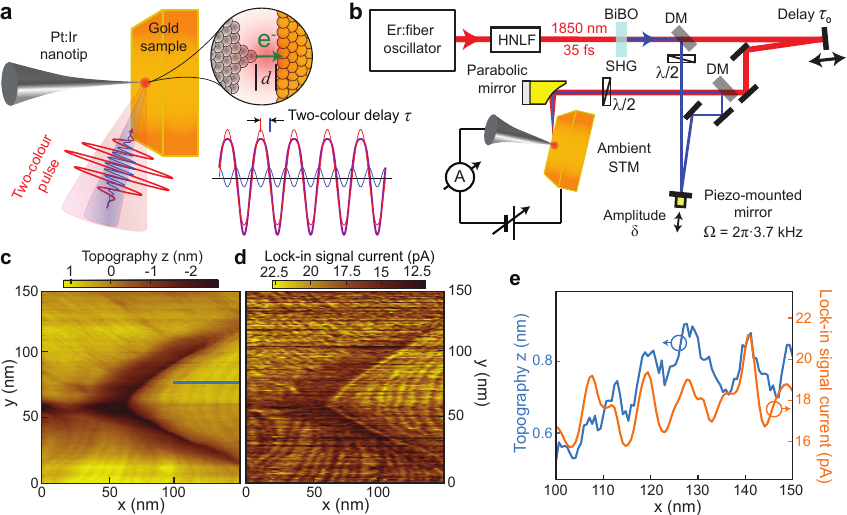}
    \caption{Ultrafast scanning tunnelling microscopy driven by two-colour laser pulses. \textbf{a}, Overview of the experiment. A Pt:Ir nanotip and a gold sample form a tunnelling junction with gap size $d$. Ultrafast currents (green) are driven by a two-colour laser pulse (red: fundamental, blue: second harmonic). Inset: The superposition of an infrared fundamental field (red) and its second harmonic (blue) leads to a field with asymmetric waveform (purple), controlled by the two-colour delay $\tau$. \textbf{b}, Optical setup of the experiment. An Er:fiber laser oscillator generates infrared pulses (red) in a highly nonlinear fiber (HNLF) assembly. Second-harmonic generation (SHG) in Bismuth Triborate (BiBO) produces the second colour (blue). We control the two-colour delay in an interferometer using a linear stage (base delay $\tau_0$) and a piezo-mounted mirror oscillating at frequency $\Omega$ (DM: dichroic mirror). The latter allows us to perform lock-in measurements of the laser-induced current without thermal artifacts. \textbf{c}, Topography obtained from a constant-current scan (sample bias 200\,mV, set-point current 100\,pA). \textbf{d}, Laser-induced lock-in signal current recorded simultaneously with the topography. The total laser power is 6\,mW and the delay $\tau_0$ is fixed. \textbf{e}, Line cross-section of the topography (blue curve) marked in c with a blue line. We also show the laser-induced lock-in signal current (orange curve). }
    \label{fig1}
\end{figure}

In our experiment, we perform STM at ambient conditions at room temperature using two-colour laser pulses (see Fig.~\ref{fig1}a). An electropolished Pt:Ir nanotip with an apex radius of $\sim$\,5...20\,nm and a gold substrate form a tunnelling junction with variable width $d$. We irradiate the junction with 35\,fs, 1850\,nm laser pulses of up to 75\,pJ pulse energy at a repetition rate of 80\,MHz together with its second harmonic (SH). The intensity ratio of the SH and the fundamental is $\sim$\,10\% in order to induce a pronounced symmetry breaking of the resulting waveform (see the inset of Fig.~\ref{fig1}a for an illustration). Figure~\ref{fig1}b shows the optical setup of our experiment (see Methods for a detailed description). We control the relative time delay between the two colours using an interferometric setup with the help of a linear delay stage. The two-colour pulses induce an optical near-field in the nano-scale tunnelling junction, leading to a strong field enhancement and enabling us to enter the strong-field sub-cycle regime of light-matter interaction.

The microscope is operated in constant-current measurement mode, where a static bias field drives a tunnelling current and a feedback loop controls the relative distance between tip and sample to achieve a constant current. In addition, ultrafast laser-induced currents are generated by the optical near-field formed inside the tunnelling junction through irradiation with the two-colour laser pulses. Isolating the contribution of the laser-induced currents from the static tunnelling current is a challenging task. A chopper-based lock-in approach (cf.~\cite{Garg2020}) has not been applicable because it produces artifact signals. Periodic on-off switching of the intensity in the kHz domain leads to expansion and contraction of the nanotip and a corresponding increase and decrease of the static tunnelling current, masking the actual laser-induced current signal (see Supplementary Information). In order to avoid any intensity modulation and the artifacts accompanying it, we modulate the two-colour delay with a sinusoidal modulation~\cite{Dolocan2011} at an angular frequency of $\Omega = 2\pi \cdot 3.7$\,kHz using a mirror mounted on a vibrating piezoelectric chip. The total two-colour delay as a function of time is then given by $\tau = \tau_0 + \delta \sin(\Omega t)$, where $\tau_0$ is the base delay set by the linear delay stage and $\delta$ is the amplitude of the sinusoidal delay modulation. This leads to a clear and consistent lock-in signal and allows us to measure the laser-induced current independently of the microscope feedback loop operating at a locking bandwidth below 500\,Hz. \REV{The two-colour delay does not affect the tip-sample distance as shown in a Fourier analysis of two-colour delay scan data (see Supplementary Information).}

Figure~\ref{fig1}c-e shows a scan of the gold sample under laser irradiation for a fixed value of $\tau_0$. The sample bias voltage is 200\,mV and the set-point current for the conventional STM operation is 100\,pA. The topography map (Fig.~\ref{fig1}\REV{c}) obtained from the static tunnelling current reveals fine atomic steps of the gold sample on the angström scale. Simultaneously with the topography, the lock-in signal at 3.7\,kHz allows us to record a map of the laser-induced current (Fig.~\ref{fig1}d), which largely follows the topographic features of the sample. The phase offset of the laser-induced current signal with respect to the delay modulation is found to be approximately uniform (see Supplementary Information). A look at a line cross-section (see Fig.~\ref{fig1}d) shows that the laser-induced current signal is sensitive to the tiny variations of the topographic height of the sample on the sub-angström scale. We find that peaks in the signal are correlated with small topographic protrusions, hinting at a remarkable sensitivity of the optical near-field to the atomic-scale sample geometry at the STM junction~\cite{Lee2019,Siday2024}. We estimate the lateral spatial resolution obtained from the laser-induced signal to be around 2\,nm, roughly the same as the spatial resolution of the conventional topographic scan. Here we are likely limited by the fact that the microscope is operated in ambient conditions.

\subsection*{Sub-cycle waveform control of tunnelling currents}

Scanning the base two-colour delay $\tau_0$ with a step size of 130\,as and placing the nanotip at a fixed spot on the sample allows us to study the sub-cycle dependence of the laser-induced current. Figure~\ref{fig2}a shows the lock-in current as a function of $\tau_0$ (see Supplementary Information for a plot of a wide-range delay scan). We find oscillations with a period of 1.5\,fs, which corresponds to half of the period of the SH field, a clear sub-cycle feature. The lock-in measurement also yields the phase of the lock-in current signal (Fig.~\ref{fig2}b). We find alternating phase jumps of 180 degrees on top of an arbitrary but constant phase offset. \REV{The lock-in signal and the laser-induced current can be related by a Fourier transform through $\mathcal{F}[I_\mathrm{lock-in}](\omega) = \mathcal{F}[I](\omega)J_1\left(\delta\omega\right)$, where $I_\mathrm{lock-in}(\tau_0)$ is the lock-in current, $I(\tau_0)$ is the laser-induced current and $J_1$ is the Bessel function of the first kind of order 1 (for the full derivation see the Supplementary Information). Reconstructing the laser-induced currents with this relation reveals that} $\tau_0$ controls the magnitude of the current, as shown by the dashed orange curve in Fig.~\ref{fig2}c; it may also indicate control of the direction of the current. However, the reconstruction from the lock-in measurements alone cannot provide any information about the offset of the current, which remains undefined \REV{since $J_1(0) = 0$}. A different, more direct measurement of the laser-induced current is required.

\begin{figure}[htb!]
    \centering
    \includegraphics[width=1\linewidth]{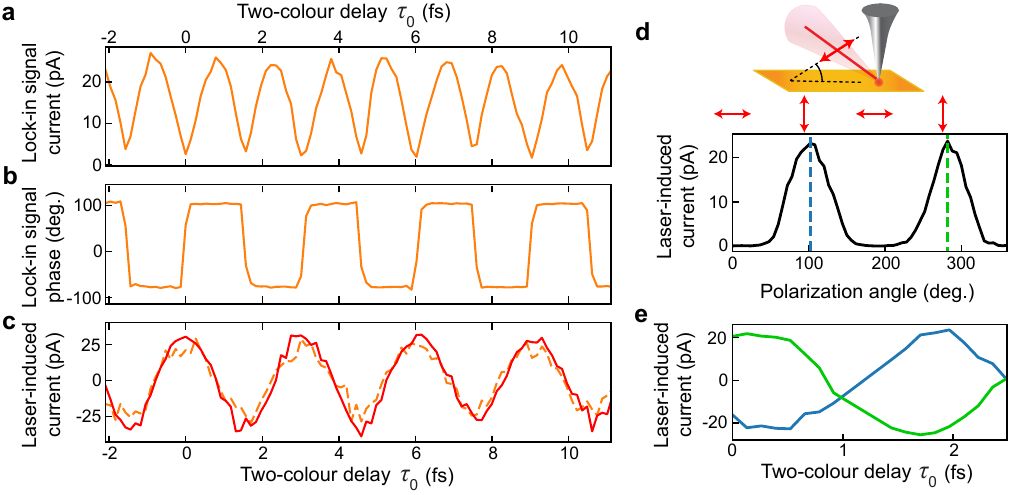}
    \caption{Laser-induced currents in the tunnelling junction. \textbf{a}, Lock-in signal current measured at a fixed position on the sample as a function of the base two-colour delay $\tau_0$  (sample bias 200\,mV, set-point current 100\,pA, 6\,mW total laser power). \textbf{b}, The corresponding lock-in phase shows periodic phase jumps of 180 degrees. \textbf{c}, Laser-induced current reconstructed from the lock-in measurement (dashed orange curve) and directly measured with frozen tip at zero bias (red curve). The direct measurement shows unambiguously that we are able to control the direction of the current. \textbf{d}, Polarization dependence of the laser-induced current measured with frozen tip and zero bias. The angle is defined with respect to the sample plane (see sketch). The peaks are marked with blue and green dashed vertical lines. \textbf{e}, Direct measurement of the laser-induced current as a function of $\tau_0$ for each of the two angles marked by the vertical lines in d.}
    \label{fig2}
\end{figure}

In order to show that $\tau_0$ indeed controls the direction of the laser-induced current, we temporarily switch off the microscope feedback loop as well as the lock-in modulation of the two-colour delay ($\delta = 0$). We freeze the position of the nanotip in space and set the sample bias to zero, thus suppressing the static tunnelling current and preempting any symmetry breaking by the static field. The only measurement quantity that remains is the net laser-induced current, now measured directly. \REV{Importantly, this approach provides the current offset that is inaccessible in the lock-in reconstruction.} The red curve in Fig.~\ref{fig2}c shows the resulting signal, which we find to be oscillatory with the period of the SH field. The signal is in good agreement with the laser-induced current reconstructed from the lock-in approach. Our measurements unambiguously demonstrate that the direction of the current is governed by $\tau_0$. Positive currents indicate that the asymmetric waveform of the two-colour field drives electrons in sub-cycle bursts from the nanotip to the sample. For negative currents, the opposite is true -- we obtain a current from the sample to the nanotip. Here, it is \DEL{solely}\REV{mainly} the laser field that breaks the symmetry of the tunnelling junction and produces a net current, a hallmark of lightwave electronics, where the sub-cycle waveform of a laser pulse governs ultrafast electric current dynamics~\cite{Borsch2023}. Since our laser pulse comprises many cycles, we obtain a train of attosecond current bursts with a periodicity of the fundamental field. \REV{CEP effects are negligible since our laser pulses are sufficiently long to avoid them (see Supplementary Information).}

Figure~\ref{fig2}d shows the \DEL{polarization }dependence of the laser-induced current \REV{on the polarization angle of the combined two-colour field} measured at zero bias while the nanotip is frozen. The polarization angle is defined with respect to the sample plane. The laser-induced current peaks sharply at a polarization angle of $\sim$\,100 degrees, which coincides approximately with the scenario in which the two-colour laser field is aligned with the perpendicularly oriented nanotip. This ensures maximum near-field enhancement in the gap and therefore maximum laser-induced current. At $\sim$\,280 degrees, the same scenario is reached, but the flip of the polarization causes a reversal of the sign of the two-colour waveform. For the two angles, we perform a two-colour delay scan and indeed observe the resulting reversal of the direction of the current (Fig.~\ref{fig2}e). This further corroborates the notion that the waveform controls the transport of electrons through the junction.

\subsection*{\REV{Transition to the strong-field regime of electron transport}}

\begin{figure}[htb!]
    \centering
    \includegraphics[width=0.7\linewidth]{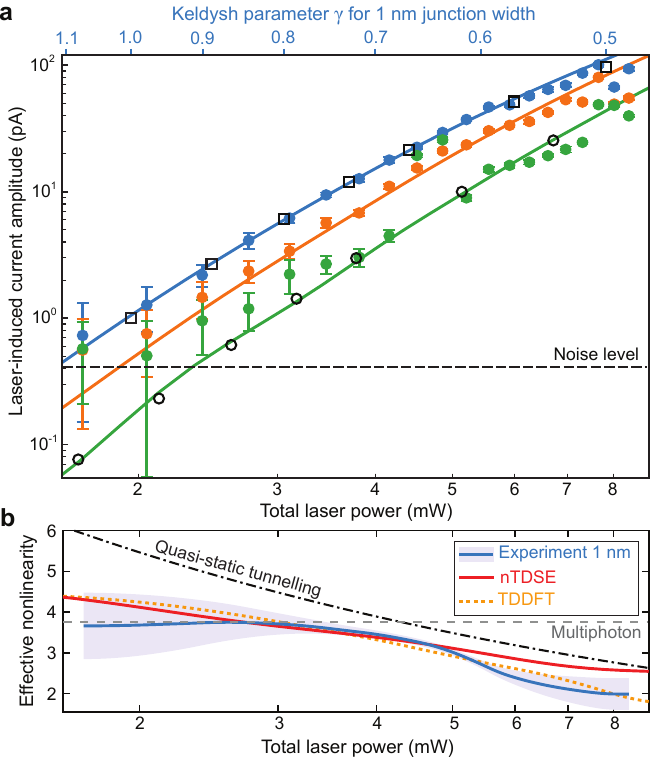}
    \caption{Power-scaling of the laser-induced current. \textbf{a}, Amplitude of the two-colour modulation of the current as a function of total laser power for three junction widths (blue: 1\,nm, orange: 1.5\,nm, green: 2\,nm; error bars: standard deviation). The solid curves display the corresponding theory curves obtained from a numerical integration of the time-dependent Schrödinger equation \REV{(nTDSE)}. \REV{In addition, we show the results of a time-dependent density functional theory (TDDFT) calculation (black open squares: 1\,nm junction width; black open circles: 2\,nm junction width). \textbf{b}, The effective power-law nonlinearity as a function of total laser power for a junction width of 1\,nm. We show the smoothed experimental data (blue solid curve with gray shaded area indicating the standard deviation) along with the nTDSE (red solid curve), the TDDFT (dotted orange curve), a multiphoton power-law with nonlinearity 3.8 (dashed gray line) and a quasi-static tunnelling curve (black dashed-dot curve).}}
    \label{fig3}
\end{figure}

A crucial question which needs to be answered is the underlying physics of the electron transport. An ultrafast STM experiment~\cite{Garg2020} and theory studies~\cite{Kim2021,Ma2024,Borisov2025,Ma2025} show that the dichotomy of a multiphoton regime and a laser-induced tunnelling regime, parametrized by the Keldysh parameter~\cite{Keldysh1965} $\gamma$, also governs ultrafast STM. In STM, the only measurement observable is the average current, unlike in nanotip photoemission experiments, for instance, where the photoelectron spectrum can reveal crucial information about the underlying physics~\cite{Kruger2011,Piglosiewicz2014,Dienstbier2023}. \REV{Thus, in the STM geometry, attosecond timing must be inferred from current measurements over control parameters such as two-colour delay, laser power, junction width and bias voltage, rather than from a directly measured electron spectrum, combined with a projection of the measurement results on theory approaches (see Supplementary Information for a detailed discussion). First,} we measure the scaling of the laser-induced current with total laser power (SH and fundamental), shown in Fig.~\ref{fig3}a in a double-logarithmic scale. Freezing the tip at zero bias, we determine the power scaling of the amplitude of the two-colour current modulation for three different junction widths, free from any other signals.  We observe that the current increases \REV{in a nonlinear way. We perform an analysis of the effective nonlinearity as a function of total laser power (Fig.~\ref{fig3}b) which shows that the current scales approximately like a power law with order 3.8 at low powers.} This may indicate the absorption of 3 or 4 photons from the two-colour pulse. However, beyond a power of \REV{3\,mW}, the nonlinearity begins to decrease. A soft kink appears in the scaling of Fig.~\ref{fig3}a, a signature of the transition to the \REV{strong-field} tunnelling regime of ultrafast STM~\cite{Garg2020,Kim2021,Ma2024}, analogous to photoemission from nanostructures~\cite{Bormann2010,Dombi2010}. Another effect that may contribute to the kink is the fact that we begin to transfer more than one electron per laser pulse starting at an average current of 13\,pA \REV{at a total laser power of 5\,mW.} 

\REV{A comprehensive theoretical analysis is needed to gain a robust insight into the underlying physics. To this end, we project our measurement data onto numerical theory models. In a growing level of complexity we employ (i) an analytical strong-field (SF) model based on the van Vleck propagator~\cite{Ma2025} allowing us to address the main physical effects;  
(ii) a single-active-electron model approach based on a numerical integration of the one-dimensional time-dependent Schrödinger equation (nTDSE), being numerically light it allows to probe the effect of the different system parameters; (iii) many-body time-dependent density function theory (TDDFT) calculations for a 2D model geometry of the junction (see Methods and Supplementary Information for details).}

\REV{First, we employ the nTDSE and directly fit the experiment with our calculations, considering the field in the junction and absolute value of electron transport as adjustable parameters (see Fig.~\ref{fig3}a). In particular, we account for the fact that the optical near-field enhancement, which translates the total laser power into the actual field driving electrons across the tunnelling junction, decreases with increasing junction width. The excellent fit to the experiment allows us to estimate the corresponding field enhancement factors for the fundamental field of $160 \pm 10$, $150 \pm 5$ and $130 \pm 10$ for the 1\,nm, 1.5\,nm and 2\,nm junction widths, respectively. These values are qualitatively in agreement with numerical solutions of the Maxwell equations for the tip-sample geometry (see Supplementary Information). We thus can also determine the effective Keldysh parameter $\gamma$ (see the Supplementary Information for the derivation of $\gamma$ for a two-colour field). The upper horizontal axis in Fig.~\ref{fig3}a shows $\gamma$ for a junction width of 1\,nm. The nTDSE model places our measurements firmly in the transition regime from multiphoton to tunnelling in the vicinity of $\gamma \sim 1$. This notion is confirmed by the TDDFT calculations which closely match the experiment in Fig.~\ref{fig3}a when considering an effective tunnelling transport area of our laser-driven STM junction of $\approx 8$\,nm$^2$ (see Supplementary Information for details) and field enhancement values of $216 \pm 40$ (1\,nm junction) and $134 \pm 30$ (2\,nm junction), close to those obtained using the Maxwell equations. The TDDFT is in good agreement with the experiment also with respect to the nonlinearity in Fig.~\ref{fig3}b.}

\begin{figure}[h!]
    \centering
    \includegraphics[width=1\linewidth]{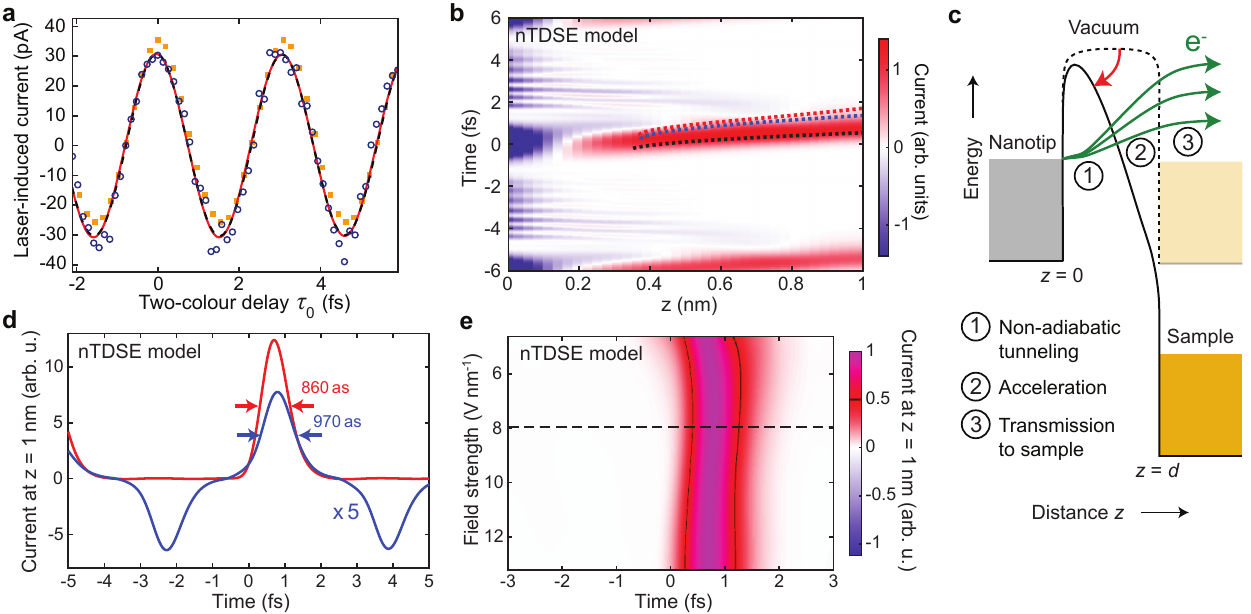}
    \caption{Tracing the attosecond dynamics in the STM junction. \textbf{a}, Laser-induced current as a function of $\tau_0$ (open blue circles, \REV{experimental} data from Fig.~\ref{fig2}c for frozen tip at zero bias). \REV{Solid red curve: nTDSE results. Dashed black curve: SF results. Filled orange squares: TDDFT results. All calculations were performed for a total peak field of $9.6\,\mathrm{V\,nm}^{-1}$ and $d = 1$\,nm.} \textbf{b}, Current density map in the tunnelling junction calculated with the nTDSE \DEL{model}with the same parameters as in a at $\tau_0 = 0$. Zero time coincides with the peak of the two-colour field. The dotted curves represent three semiclassical electron trajectories obtained from the SF model (from bottom to top: final kinetic energies 6.7\,eV, 4.4\,eV and 0\,eV with respect to the vacuum level). The trajectories ``emerge'' from the tunnel exit located at $z \sim 0.35$\,nm. \textbf{c}, Illustration of the three-step model of ultrafast STM in the non-adiabatic tunnelling regime. Here the electron gains energy already under the tunnelling barrier \REV{thinned by the light field and is then further accelerated towards the sample}. \textbf{d}, Current density at $z = 1$\,nm obtained from the nTDSE \REV{calculations} for the two-colour pulse (red, strong symmetry breaking) and a single-colour pulse (blue, no symmetry breaking, multiplied by a factor of 5). We also mark the corresponding burst durations. \textbf{e}, Normalized current density at $z = 1$\,nm \REV{calculated with the nTDSE} for the two-colour pulse as a function of peak field strength. The dashed line marks $\gamma = 1$ \REV{and the time measured between black curves for the fixed field indicates the FWHM duration of the current burst}. }
    \label{fig4}
\end{figure}

\REV{The sub-cycle waveform control of the direction of the attosecond current bursts can also be explained with the help of the three theory approaches (SF, nTDSE, TDDFT). Figure~\ref{fig4}a shows the measured laser-induced current from Fig.~\ref{fig2}c. In addition, we display the curves obtained from the three theory approaches for a total peak field strength of $9.6\,\mathrm{V\,nm}^{-1}$ ($\gamma \sim 0.8$) after scaling their amplitude to match the experimental data. We find good agreement. The experimentally observed current modulation is nearly completely symmetric with respect to zero current, but not perfectly; a slight asymmetry is observed in the experimental data and in the TDDFT results. An analysis of the TDDFT calculations shows that, e.g., geometrical and work function differences between tip and sample affect the magnitude and direction of the asymmetry (see Supplementary Information).}

\subsection*{\REV{Reconstruction of the dynamics of} the attosecond current bursts}

The excellent overall agreement between experiment and theory allows us to trace the electron dynamics and estimate the duration of the attosecond electron bursts \REV{from the calculations}. Figure~\ref{fig4}b shows the current density map calculated with the nTDSE as a function of space and time across a 1\,nm junction for a \REV{total} peak field strength of $8\,\mathrm{V\,nm}^{-1}$ (\REV{$\gamma \sim 0.96$}) and $\tau_0 = 0$ for maximum symmetry breaking. \REV{In agreement with the TDDFT results (Supplementary Material), the nTDSE results indicate that t}he electrons are created at the peak of the field (time zero) and cross the \REV{junction} on a scale much shorter than the optical cycle duration and hence without any quiver motion~\cite{Ludwig2020,Ma2024,Borisov2025}. The flow of the current density \REV{and the acceleration} match\DEL{es} well with semiclassical trajectories obtained from the SF model that appear after the tunnelling exit located at $\sim$\,0.35\,nm (see dotted curves in Fig.~\ref{fig4}b). \REV{All three theoretical approaches (SF, nTDSE, TDDFT) point to the three-step model of ultrafast STM in the non-adiabatic tunnelling regime: Tunnelling through a barrier thinned by the strong light field and concomitant energy gain (step 1), further acceleration inside the junction (step 2) and transmission into the sample (see Fig.~\ref{fig4}c for an illustration).} The full-width at half maximum (FWHM) duration of the electron burst at the end of the junction at $z = 1$\,nm \REV{as derived from the nTDSE} is depicted in Fig.~\ref{fig4}d and amounts to \REV{$860 \pm 90$\,as, which accounts for uncertainties related to experimental parameters}. \REV{This result is further supported by quantitative agreement between nTDSE and TDDFT in the description of the electron current dynamics in the junction as discussed in the Supplementary Information. It is also} shorter than 970\,as obtained in the single-colour case (1850\,nm field only), showing that the two-colour approach not only induces full symmetry breaking, but also leads to a shorter burst duration within the sub-femtosecond time domain. \REV{From the nTDSE results shown in} Fig.~\ref{fig4}\REV{d}, we can also \REV{deduce} that the \REV{calculated current burst} arrival time is about 700\,as after the peak of the field, which accounts for the travel time through the junction \REV{and should not be confused with a strong-field tunnelling time~\cite{Eckle2008,Sainadh2019}}. The burst duration is strongly confined to the sub-cycle regime for a wider range of peak field strengths (see Fig.~\ref{fig4}\REV{e}), showing that attosecond current bursts can be generated routinely. The same is true for variations of more experimental parameters, such as the intensity ratio of the SH and the fundamental, junction width and tip workfunction; the burst duration stays well below the 1\,fs mark (see Supplementary Information). While our experiment is performed at an intensity ratio of the SH and the fundamental of $\sim$\,10\%, ratios as small as 2.5\% can induce strong symmetry breaking (see Supplementary Information).

\REV{Looking ahead, replacing two-colour pulses with single-cycle drivers will confine electron transport to a single isolated attosecond bursts~\cite{Rossetti2025}. Nevertheless, two-colour pulses remain attractive for probing ultrafast charge transport in STM: they are easy to generate, allow tunable and strong waveform asymmetry for current reversal, and yield larger current signals due to their multi-cycle nature. A possible route to shorter current bursts is reducing the driver wavelength, but this is limited by the need to maintain sufficient nonlinearity, set by the ratio of photon energy and work function. Alternatively, stronger fields can broaden the transmitted electron energy spectrum and, in principle, shorten the bursts~\cite{Ma2024,Borisov2025}. In practice, however, electron dispersion across the finite junction keeps the burst duration nearly constant below $\gamma \sim 1$~\cite{Ma2024}. Broader spectra also smear the electron energies and ultimately reduce energy resolution, which may limit future applications.}

\section*{Discussion}

\REV{Our work directly demonstrates two-colour sub-cycle waveform-controlled reversal of laser-driven current in an STM junction. It further shows that such waveform control can induce a train of sub-femtosecond current bursts across a variable STM junction while maintaining a lateral spatial resolution of 2\,nm and sensitivity to sub-ångström topographic features under ambient conditions and free from thermal artifacts.} In the future, our approach can be combined with all-optical readout~\cite{Siday2024} to directly reveal the evolution of the sub-cycle currents. Moving from a two-colour laser pulse to a single-cycle infrared pulse will naturally lead to a single isolated attosecond electron burst with waveform-controlled timing and direction. Combining the ability to generate these bursts with the well-established capability to image electronic states in molecules, defects, and nanostructures using conventional STM in ultrahigh vacuum conditions, we envision the combined real-time and real-space observations of coherent electron-hole dynamics and many-body effects as they unfold. For example, a sub-cycle controlled infrared pump pulse will enable extracting an electron from a molecule with attosecond precision in time at a specific atomic site, inducing a coherent evolution of complex electron-hole dynamics. A second, sub-cycle controlled probe pulse with opposite field waveform can then inject an electron back into the molecule, allowing for a readout of the dynamics before dephasing effects set in. With attosecond STM, microscopy has reached its ultimate spatiotemporal resolution limits at the atomic scale.

\section*{Methods}

\subsection*{Experimental setup}

100\,fs laser pulses from an Er:fiber laser system (Menlo C-Fiber High Power, 500\,mW, 1550\,nm, 80\,MHz\REV{, CEP is not stabilized}) are focused into a fiber assembly (standard polarization-maintaining fiber of 75\,mm length spliced to 12\,mm of highly nonlinear fiber, Thorlabs HN1550P) for supercontinuum generation spanning from 950\,nm to 2100\,nm (cf.~\cite{Brida2014}). We use only the solitonic part of the resulting supercontinuum with a central wavelength of 1850\,nm by rejecting the spectral components below 1500\,nm using a dichroic filter. These pulses are focused into a bismuth triborate (BiBO) crystal with a thickness of 1\,mm for second harmonic (SH) generation, avoiding the messy part of the spectrum below 1650\,nm by an appropriate choice of phase-matching conditions. The recollimated fundamental and SH beams are then fed into an interferometric setup where they are separated by a dispersion-engineered dichroic mirror. The SH beam passes through several filters to remove any remaining trace of the fundamental beam. Subsequently, its polarization is rotated to match the polarization of the fundamental. It then hits a mirror glued onto a piezo chip at a near-zero angle of incidence. The piezo can be modulated at kHz frequencies using a high-voltage signal to periodically change the temporal delay of the SH beam, providing the modulation to produce the lock-in signal. The delay modulation amplitude in our experiment is $\delta = 0.6$\,fs. The fundamental beam is passed through 1\,mm of silicon to compensate for the dispersion of the other optical elements in the system and a longpass filter at 1650\,nm to remove the messy part of the spectrum below that wavelength. We also control the base two-colour delay $\tau_0$ using a mirror mounted on a linear closed-loop piezo stage at a near-zero angle of incidence. Both beams are recombined by another dispersion-engineered dichroic mirror and expanded by a factor of 3. We monitor the two-colour delay using an auxiliary collinear interferometer setup and correct for drifts. 

Before sending the beams to the microscope (RHK Technologies PanScan Flow Kit operated in ambient conditions), we control the laser power of the combined beams using a variable neutral-density filter and tune their joint polarization by an achromatic half-wave plate. An off-axis parabolic mirror with a focal length of 50\,mm focuses the beams on the tip-sample junction at an angle of incidence of 20 degrees with respect to the sample plane. We use commercial Pt:Ir nanotips (Unisoku P-100PtIr(S)) with apex curvature radii of 5...20\,nm and a flat gold substrate fabricated according to a self-formation procedure~\cite{Borukhin2012}. The fundamental laser beam is focused to a spot size of $\sim$\,17\,$\mu$m. The duration of the pulse is 35\,fs. The SH beam is focused to a spot size of $\sim$\,10\,$\mu$m. The duration of the SH pulse is 62\,fs due to uncompensated chirp. Taking into account all parameters \REV{including field enhancement and chirp}, the ratio of the SH intensity to the fundamental intensity is about 10\,\%. The gold sample can be biased, and the current is measured using a current preamplifier connected to the tip. In the tip freezing experiments, we switch off the bias and the microscope feedback loop for a duration of $\sim$\,600\,ms and take several data points during this time. 

\subsection*{Theory \REV{approaches}}

\REV{We employ three theory approaches in our work, (i) an analytical strong-field (SF) model based on the van Vleck propagator~\cite{Ma2025};  (ii) a numerical integration of the time-dependent Schrödinger equation (nTDSE); (iii) the many-body time-dependent density functional theory (TDDFT) for a 2D model geometry of the STM junction. The first two models are based on strong approximations because they consider only a single active electron and the system is reduced to only one dimension. Despite the high degree of simplification, these theory approaches have} yielded excellent agreement with nanotip experiments~\cite{Kruger2011,Dienstbier2023}, for instance. \REV{In our case, tip} and sample are \REV{modelled} as potential wells and populated with a wavefunction at the Fermi level\DEL{(see Supplementary Information for more details)}. In the following, we briefly describe the SF model~\cite{Ma2024,Ma2025} \REV{and refer the reader to the Supplementary Information for details of the nTDSE and TDDFT}.

Based on the Dyson equation, we define the tunnelling amplitude $M_{E}$ from tip to sample as
\begin{equation}
\begin{split}
M_{E}= \frac{i\hbar}{2m}\int_{-\infty}^{\infty} \bigg[\Psi_{\mathrm{Is}}(z,t)\frac{\partial}{\partial z}\psi^{*}(z,t)
-\,\psi^{*}(z,t)\frac{\partial}{\partial z}\Psi_{\mathrm{Is}}(z,t)\bigg]\bigg\vert^{z=d}_{z=0}\;dt,
\label{S2.1}
\end{split}
\end{equation}
where $\Psi_\mathrm{Is}(z,t)$ and $\psi^{*}(z,t)$ are time-dependent wavefunctions in the junction and the sample, respectively. The notation $[...]\big\vert^{z=d}_{z=0}$ at the end of Eq.~\ref{S2.1} indicates the subtraction of the term inside the brackets evaluated at $z=0$ from the term evaluated at $z=d$. 

In analogy to the strong-field approximation of attosecond science, the wavefunctions $\Psi_\mathrm{Is}(z,t)$ and $\psi^{*}(z,t)$ can be obtained by using the Van Vleck propagator and the eigenfunction of the sample system~\cite{Ma2025}. The simplified tunnelling amplitude is now 
\begin{eqnarray}\label{S2.2}
M_E&=&\int_{-\infty}^{\infty}\int_{-\infty}^{t_{2}}\sqrt{\frac{i}{8\pi m\hbar^3(t_{2}-t_{1})}}
\eta(t_{1}, t_{2}) e^{\frac{i}{\hbar}S(t_{2},t_{1})}\;dt_{1}dt_{2}\,.
\end{eqnarray}
Here, $\eta(t_{1}, t_{2})$ is a prefactor originating from the transition matrix elements. For an initial bound state $E_{0}=-|E_{0}|$ in the tip and a final state $E$ in the sample, the action $S(t_{2},t_{1})$ in the exponent is given by
\begin{eqnarray}\label{S2.3}
S(t_{2},t_{1})&=&E t_{2}+\frac{{\tilde p}^2}{2m}(t_{2}-t_{1})-\int_{t_{1}}^{t{2}}\frac{e^2A^2(\tau)}{2m} \;d\tau\,\nonumber\\
&&-\int_{t_{1}}^{t{2}}V_{\mathrm{imag}}[z(\tau)]\;d\tau\,+\vert E_{0}\vert t_{1},
\end{eqnarray}
where $\tilde p=\frac{\int_{t_{1}}^{t_{2}}{eA(\tau)}\;d\tau\,+md}{t_{2}-t_{1}}$ is the effective canonical momentum, $A(\tau)$ is the vector potential of the laser field, $d$ is the junction width, and $V_{\mathrm{imag}}$ is the image potential. We use $t_{1}$ and $t_{2}$ to represent the emission time and the arrival time of the electron transport from tip to sample.

Following the saddle-point technique with $\nabla_{t_{1},t_{2}}S(t_{2},t_{1})=0$, we can obtain the three saddle-point equations:
\begin{eqnarray}
\frac{\left[\tilde p_\mathrm{(s)}-eA(t_{\mathrm{1s}})\right]^2}{2m}-\overline{|V_{\mathrm{imag}}|}=-\left\vert E_0\right\vert,\label{S2.4}
\\
\int_{t_\mathrm{1s}}^{t_\mathrm{2s}}\frac{\left[\tilde p_\mathrm{(s)}-eA(\tau)\right]}{m}\;d\tau\,=d,\label{S2.5}
\\
\frac{\left[\tilde p_\mathrm{(s)}-eA(t_\mathrm{2s})\right]^2}{2m}-\overline{|V_{\mathrm{imag}}|}=E.\label{S2.6}
\end{eqnarray}
Here, the subscript (s) indicates that these values correspond to complex-valued mathematical saddle points rather than physical quantities. $\overline{|V_{\mathrm{imag}}|}$ is the image potential averaged over the length of the junction. The three equations above provide a three-step framework for describing electron transport.
(1) At the moment of emission, the bound electron is released from the tip by the laser field. This process is governed by the energy condition in Eq.~\ref{S2.4}. 
(2) After emission, the electron is accelerated by the laser field as it travels from the tip toward the sample. Its motion is described by Eq.~\ref{S2.5}. Unlike the characteristic recollision dynamics in attosecond science, the electron does not return to its origin.
(3) Transmission into the sample: The energy accumulated during the second step is transferred to the sample, as described by Eq.~\ref{S2.6}. In Newtonian mechanics, the displacement of a classical point-like electron in the laser field is 
\begin{equation}\label{S2.7}
{\cal D}(t)=\int_{t_\mathrm{1s}}^{t}\frac{\left[\tilde p_{(\mathrm{s})}-eA(\tau)\right]}{m}\;d\tau\,.
\end{equation}
Since the saddle points are complex, the integral must be taken also over a complex contour. In order to display the resulting semiclassical trajectories in Fig.~\ref{fig4}b, we take the real part of ${\cal D}(t)$.

\section*{Data availability}

The data that support the plots within this paper and other findings of this study are available from the authors upon request.

\section*{Code availability}

The codes that support the findings of this study are available from the authors upon request.

\section*{Author contributions}

D.~D., B.~M.~and A.~G.~contributed equally to this work.

M.~K.~conceived and supervised the project. A.~G., S.~C., B.~M.~and M.~K.~designed and built the experimental setup. D.~D., A.~G., B.~M.~and M.~K.~performed the measurements. B.~M.~carried out the \REV{nTDSE and SF model} calculations. \REV{A.~G.~B. devised and performed the TDDFT calculations.} M.~K.~wrote the initial manuscript. All authors contributed to the preparation of the final manuscript.

\section*{Corresponding authors}

Correspondence to \href{mailto:krueger@technion.ac.il}{Michael Kr\"uger}.

\section*{Acknowledgements}

The authors thank M.~Ivanov, A.~Kollin and Y. Bekenstein for insightful discussions, I.~Kaminer for providing lab space for initial experiments, A.~Feigenboim for supporting the design of experimental system, I.~Sthzeglowski for the initial setup of the fiber laser, P. Sidorenko for assistance with fiber splicing, and  N.~Dudovich, U.~Leonhardt and L.~M.~Procopio for providing specialized equipment. \REV{A.~G.~B.~gratefully acknowledges the warm hospitality of DIPC.} 

\section*{Funding statement}

This project has received funding from the European Union's Horizon 2020 research and innovation program under grant agreement No 853393-ERC-ATTIDA and from the Israel Science Foundation (ISF) under grant agreement No 1504/20. We also acknowledge the Helen Diller Quantum Center and the Russell Berrie Nanotechnology Institute at the Technion for partial financial support.

\section*{Ethics declarations}

The authors declare no competing interests.


\end{document}


\title{Supplementary Information: \\
Attosecond current control and timing in a scanning tunnelling microscope}

\author{Daniel Davidovich}
\altaffiliation{These authors contributed equally to this work.}
\affiliation{Department of Physics, Technion---Israel Institute of Technology, Haifa 32000, Israel}
\affiliation{Solid State Institute, Technion---Israel Institute of Technology, Haifa 32000, Israel}
\affiliation{The Helen Diller Quantum Center, Technion---Israel Institute of Technology, Haifa 32000, Israel}

\author{Boyang Ma}
\altaffiliation{These authors contributed equally to this work.}
\affiliation{Department of Physics, Technion---Israel Institute of Technology, Haifa 32000, Israel}
\affiliation{Solid State Institute, Technion---Israel Institute of Technology, Haifa 32000, Israel}
\affiliation{The Helen Diller Quantum Center, Technion---Israel Institute of Technology, Haifa 32000, Israel}

\author{Adi Goldner}
\altaffiliation{These authors contributed equally to this work.}
\affiliation{Department of Physics, Technion---Israel Institute of Technology, Haifa 32000, Israel}
\affiliation{Solid State Institute, Technion---Israel Institute of Technology, Haifa 32000, Israel}
\affiliation{The Helen Diller Quantum Center, Technion---Israel Institute of Technology, Haifa 32000, Israel}

\author{Shimon Cohen}
\affiliation{Department of Physics, Technion---Israel Institute of Technology, Haifa 32000, Israel}
\affiliation{Solid State Institute, Technion---Israel Institute of Technology, Haifa 32000, Israel}
\affiliation{The Helen Diller Quantum Center, Technion---Israel Institute of Technology, Haifa 32000, Israel}

\author{Zhaopin Chen}
\affiliation{Department of Physics, Technion---Israel Institute of Technology, Haifa 32000, Israel}
\affiliation{Solid State Institute, Technion---Israel Institute of Technology, Haifa 32000, Israel}
\affiliation{The Helen Diller Quantum Center, Technion---Israel Institute of Technology, Haifa 32000, Israel}

\author{\REV{Andrei G. Borisov}}
 \affiliation{Institut des Sciences Mol\'eculaires d'Orsay (ISMO), UMR 8214, CNRS, Universit\'e Paris-Saclay, 91405 Orsay Cedex, France.}
      \affiliation{Donostia International Physics Center, Manuel de Lardizabal 4, 20018 Donostia, Spain}

\author{Michael Kr\"uger}
\altaffiliation{Corresponding author: krueger@technion.ac.il}
\affiliation{Department of Physics, Technion---Israel Institute of Technology, Haifa 32000, Israel}
\affiliation{Solid State Institute, Technion---Israel Institute of Technology, Haifa 32000, Israel}
\affiliation{The Helen Diller Quantum Center, Technion---Israel Institute of Technology, Haifa 32000, Israel}

\date{\today}

\keywords{}

\maketitle

\section{Thermal artifacts in chopper-based lock-in measurements}

\begin{figure}[hb!]
    \centering
    \includegraphics[width=0.55\linewidth]{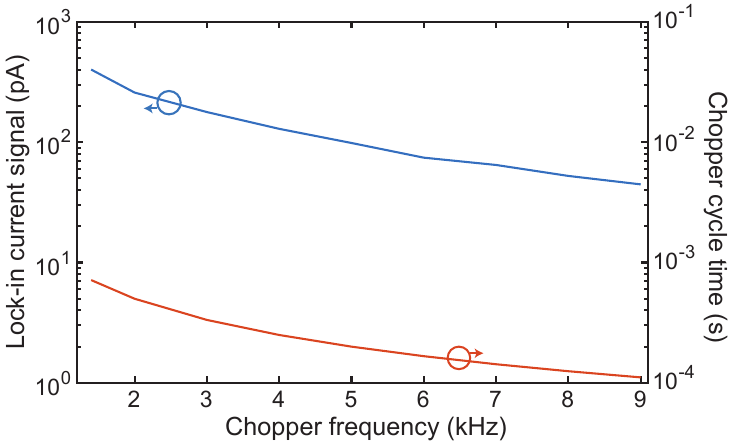}
    \caption{Thermal artifacts from chopping the laser beam. Lock-in current signal (blue) and chopper cycle time (orange) as a function of chopper frequency (single colour only, 6\,mW laser power, sample bias voltage 500\,mV, set-point current 500\,pA).}
    \label{figChop}
\end{figure}

Periodic chopping of a laser beam is a standard approach to isolate a laser-induced signal from other signals, which has also been applied to ultrafast STM~\cite{Garg2020}. Here, we make a simple test to check the performance of this approach and chop single-colour laser pulses incident on the tunnelling junction at an average power of 6\,mW with variable chopper frequency (see Fig.~\ref{figChop}). A well-behaved measurement would reveal a lock-in current signal which does not depend on the chopper frequency. However, we observe that the lock-in current signal strongly depends on the frequency, indicating that the measured signal is influenced by the chopper. We also plot the chopper cycle duration in a logarithmic scale and find that the lock-in current signal follows the behaviour of the chopper cycle duration. This is a clear signature of a thermal heating effect that depends on how much time we give the tip to expand or contract. The chopping approach is unsuitable for laser-induced current measurements in our experiments.

\section{Influence of two-colour delay on tip-sample distance}

\begin{figure}[htb!]
    \centering
    \includegraphics[width=\linewidth]{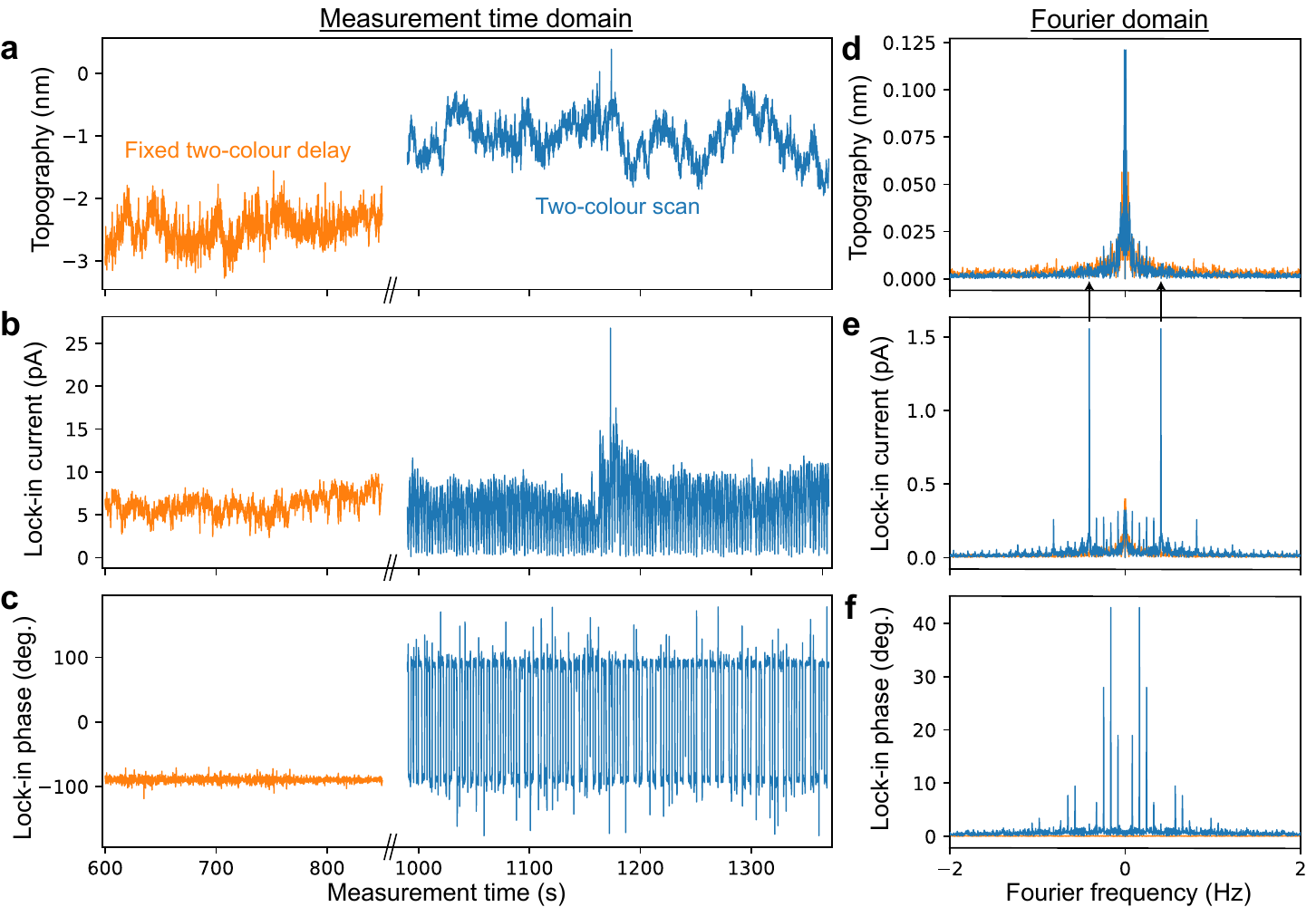}
    \caption{Fourier analysis of lock-in STM measurements. \textbf{a}, \textbf{b}, and \textbf{c} show topographic height, lock-in current and lock-in phase, respectively, for a measurement with fixed delay (orange) and scanned delay (blue). \textbf{d}, \textbf{e}, and \textbf{f} display the Fourier amplitude spectrum for those quantities. The black arrows mark the Fourier components which are related to the two-colour modulation of the lock-in current. These components do not show up in the Fourier amplitude of the topography.}
    \label{fig:Fourier}
\end{figure}

Our lock-in approach is based on the fact that the two-colour delay leads to a strong modulation of the laser-induced current in the STM junction without any modulation to the laser power in order to avoid thermal effects altogether. However, it is indeed conceivable that there may be an undesired effect of the peak electric field on the tip-sample distance. This quantity crucially depends on the two-colour delay. In order to quantify this potential effect causing an artifact signal, we performed an analysis of a two-colour delay scan measured using the lock-in technique at otherwise fixed experimental parameters (total laser power 5\,mW, fixed position on the sample, 200\,mV sample bias voltage, 100\,pA set-point current) and monitoring the topographic height kept by the feedback loop, which is a tell-tale sign for a change in the tip-sample distance. In a Fourier analysis with respect to the measurement time axis, we see no detectable correlation between the characteristic topographic variation and the lock-in current modulation above the background noise which carries a Fourier amplitude of 10\,pm (see Fig.~\ref{fig:Fourier}).

\section{Lock-in signal phase}

Here we show the phase corresponding to Fig.~1d in the main text. It is almost completely flat around a constant offset value of 7 degrees.

\begin{figure}[hb!]
    \centering
    \includegraphics[width=0.5\linewidth]{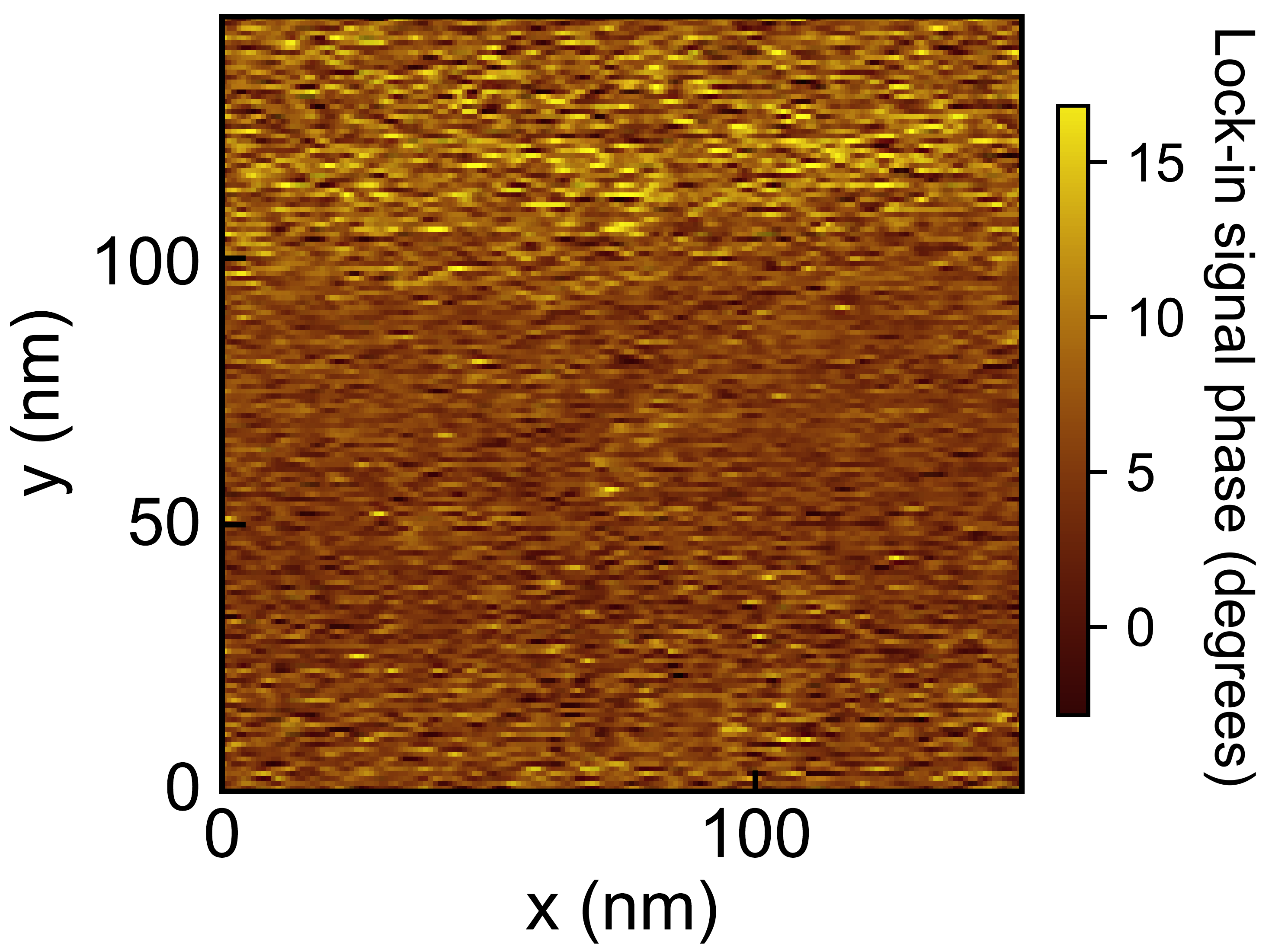}
    \caption{Lock-in signal phase corresponding to Fig.~1d in the main text (6\,mW total laser power, sample bias voltage 200\,mV, set-point current 100\,pA).}
    \label{figPhase}
\end{figure}

\section{Long-range two-colour delay scan}

\begin{figure}[htb!]
    \centering
    \includegraphics[width=0.85\linewidth]{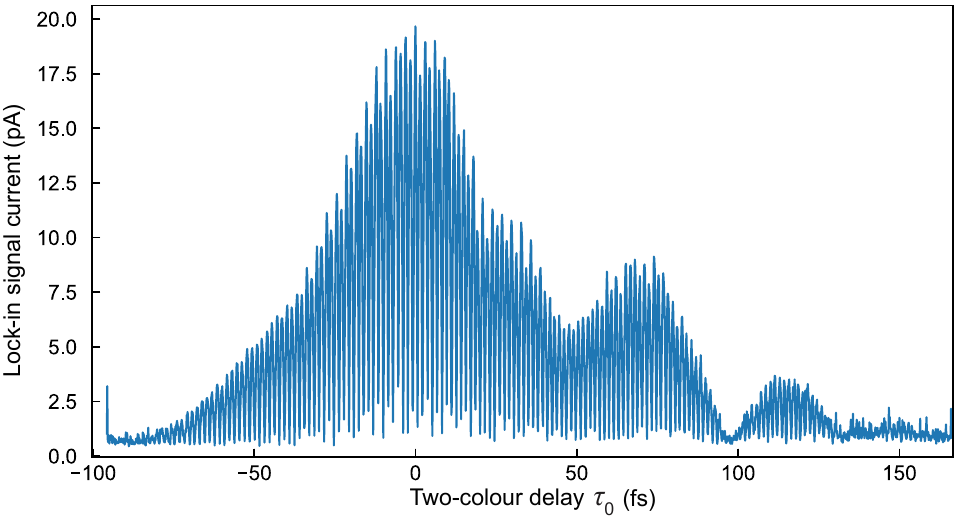}
    \caption{Long two-colour delay scan (6\,mW total laser power, sample bias voltage 200\,mV, set-point current 100\,pA).}
    \label{figLong}
\end{figure}

Figure~\ref{figLong} displays a long-range two-colour delay scan using the lock-in approach in the main text. We find that the signal persists over a range of 100\,fs around the main peak, with two smaller peaks appearing at positive delays $\tau_0 \sim 75$\,fs and $\tau_0 \sim 115$\,fs. We attribute these features to the fact that the SH pulse is fairly long and its chirp remains largely uncompensated. Also we note that the symmetry breaking and hence the laser-induced current is not very sensitive to the intensity ratio SH/fundamental, causing the two-colour oscillations to be spread over a wide range of delays.

\section{Reconstruction of the laser-induced current from the lock-in signal}\label{sec3}

In our lock-in experiment, we modulate the two-colour time delay at an angular frequency $\Omega$ and amplitude $\delta$ and use a lock-in amplifier to measure the laser-induced current. Given that the laser-induced current for a base two-colour delay $\tau_0$ between the two beams is given by $I(\tau_0)$, the lock-in current signal for delay $\tau_0$ is
\begin{equation}\label{SR1}
    I_\mathrm{lock-in}(\tau_0) = \frac{\Omega}{2\pi} \int_{-\frac{\pi}{\Omega}}^{\frac{\pi}{\Omega}}I\left(\tau_0 + \delta\sin(\Omega t)\right)e^{-i\Omega t}dt = \frac{1}{2\pi}\int_{-\pi}^\pi I\left(\tau_0 + \delta\sin(x)\right)e^{-ix}dx.
\end{equation}
Assuming that $I_\mathrm{lock-in}(\tau_0)$ describes the measurement, we can apply the Fourier transform
\begin{eqnarray}\label{SR2}
    \mathcal{F}[I_\mathrm{lock-in}](\omega) &=& \frac{1}{2\pi}\int_{-\infty}^\infty \int_{-\pi}^\pi I\left(\tau_0 + \delta\sin(x)\right)e^{-ix - i\omega\tau_0}dxd\tau_0 \nonumber\\
    &=& \frac{1}{2\pi}\int_{-\infty}^\infty \int_{-\pi}^\pi I\left(u\right)e^{-ix - i\omega(u - \delta\sin(x))}dxdu \\
&=& \mathcal{F}[I](\omega)J_1\left(\delta\omega\right).\nonumber
\end{eqnarray}
Here, $J_1$ is the Bessel function of the first kind of order 1, and $\mathcal{F}$ is the Fourier transform. To deal with the zeros of $J_1$, we introduce a cutoff $\beta$ and define
\begin{eqnarray}\label{SR3}
J_1^{(\beta)}(x) = \begin{cases}
J_1(x) & |J_1(x)| > \beta, \\
\frac{J_1(x)}{|J_1(x)|}\beta & 0 < |J_1(x)| \le \beta, \\
\infty & J_1(x) = 0.
\end{cases}
\end{eqnarray}
Then, to reconstruct $I$ from $I_\mathrm{lock-in}$ we use the formula:
\begin{equation}\label{SR4}
I(\tau_0) \approx \mathcal{F}^{-1}\left[\mathcal{F}[I_\mathrm{lock-in}](\omega) / J_1^{(\beta)}\left(\delta\omega\right)\right].
\end{equation}
As in $J_1(0)=0$, the absolute offset of the current cannot be reconstructed and therefore the direction of the current is unknown. $\beta$ is chosen to be as small as possible until it is so small that the noise begins to be amplified too much (due to the division in Eq.~\ref{SR4}). The results are shown in Fig.~2c of the main text.

\section{Time-dependent Schr\"odinger equation description of ultrafast STM}\label{secTDSE}

In the one-dimensional case, the general form of the time-dependent Schr\"odinger equation (TDSE) is given by:
\begin{equation}\label{ST1}
i\hbar\frac{\partial}{\partial t}\Psi(z,t)=-\frac{\hbar^2}{2m}\frac{\partial^2}{\partial z^2}\Psi(z,t)+V(z,t)\Psi(z,t).
\end{equation}
The potential $V(z,t)$ consists of a static junction potential and the laser interaction. Under the length gauge, the static potential of the STM is described as:
\begin{equation}\label{ST2} V_{0}(z) = \begin{cases} -(E_\mathrm{F, t}+W_\mathrm{t}), &  x < 0 ,\\ V_\mathrm{imag}(z)-e(\varphi+U_\mathrm{s}) z/d, & 0 \leqslant z \leqslant d,\\ -(E_\mathrm{F, s}+W_\mathrm{s}+e\varphi+e U_\mathrm{s}), & z > d, \end{cases}
\end{equation}
where $e = -|e|$ is the charge of the electron and $V_\mathrm{imag}(z)$ is the image potential in the junction. $\varphi=(W_\mathrm{t}-W_\mathrm{s})/e$ is the contact potential difference, which is also known as Volta potential, and $U_\mathrm{s}$ is the static bias voltage applied to the junction. We set the tip boundary at $z=0$ and the sample boundary at $z=d$. According to the Simmons theory, the image potential inside the gap is described as~\cite{Yoshioka2016, Simmons1964}
\begin{equation}\label{ST3}
V_\mathrm{imag}(z)=\left(-\frac{e^2}{8\pi\varepsilon}\right)\bigg[\frac{1}{2z}+\sum^{\infty}_{n=1}{\left\{\frac{nd}{(nd)^2-z^2}-\frac{1}{nd}\right\}}\bigg].
\end{equation}
Here, the permittivity $\varepsilon$ of the vacuum gap is $1$. The singularities of the image potential are removed by truncating the potential at the depth of each potential well, as noted in the original Simmons' paper~\cite{Simmons1964}.

In the interaction, the tip and sample are considered ideal metals, ensuring that the laser field is perfectly screened. Here we only consider the case where the polarization of the laser electric field is along the $z$-axis, so the interaction is given by
\begin{equation}\label{ST4} 
V_\mathrm{I}(z,t) = \begin{cases} 0, &  z < 0 ,\\ -e{\cal E}(t)z, & 0 \leqslant z \leqslant d,\\ -e{\cal E}(t)d, & z > d, \end{cases} \end{equation}
where ${\cal E}(t)=-\frac{\partial A(t)}{\partial t}$ is the electric field of the laser. $A(t)$ is the vector potential with the definition
\begin{equation}\label{ST5} 
A(t) =\frac{F_1}{\omega} \exp{\left[-\frac{2\ln(2)}{\tau_{1}^2}t^2\right]}\sin(\omega t)+\frac{F_2}{2\omega} \exp{\left[-\frac{2\ln(2)}{\tau_{2}^2}\left(t-\frac{\phi}{2\omega}\right)^2\right]}\sin(2\omega t-\phi). \end{equation}
The pulse duration $\tau_{1}$ (fundamental) as well as $\tau_{2}$ (second harmonic) are defined as the \REV{full width at half maximum (FWHM) of the intensity envelope}, $F_1$ and $F_2$ are the effective field strengths after near-field enhancement, $\omega$ is the angular frequency of the central wavelength, and $\phi$ is the relative phase.
The dipole approximation is justified here because the width of the vacuum junction is much smaller than the laser wavelength~\cite{Garg2021}.

In our calculations, the Fermi energies of both the Pt-Ir tip and the gold sample are approximately equal, with $E_\mathrm{F,t}\approx E_{\mathrm{F,s}}\approx 5$ eV. Since the workfunctions of Au (5.1–5.4 \,eV), Pt (5.1–5.9 \,eV), and Ir (5.0–5.7 \,eV) are also similar,  we use $W_{\mathrm{t}}\approx W_{\mathrm{s}}\approx 5.1$\,eV. The pulse durations of the fundamental and second harmonic beams are $\tau_1=35$ fs and \REV{$\tau_2=62$
fs}, respectively. The ratio of their field strengths is given by $\eta=\frac{F_2}{F_1}=\sqrt{0.1}=0.32$.

\section{Numerical integration of the TDSE (\lowercase{n}TDSE)}\label{sec1}

We solve the TDSE numerically using the Crank–Nicolson method. If the wavefunction at the initial time $t_0$ is known as $\Psi(z,t_0)$, then the propagated wavefunction at the next infinitesimal time $t_0+dt$ is
\begin{equation}\label{S1.2}
\Psi(z,t_0+dt)=\mathrm{exp}[-\frac{i}{\hbar}H(z,t_0)dt]\Psi(z,t_0).
\end{equation}
Consequently, the wavefunction at an arbitrary future time $t$ can be expressed as the cumulative evolution:
\begin{eqnarray}\label{S1.3}
\Psi(z,t)&=&\mathrm{exp}[-\frac{i}{\hbar}H(z,t-dt)dt]...\mathrm{exp}[-\frac{i}{\hbar}H(z,t_0)dt]\Psi(z,t_0)\nonumber\\
&=&\mathrm{exp}[-\frac{i}{\hbar}H(z,t-dt)dt]\Psi(z,t-dt).
\end{eqnarray}
Therefore, solving the TDSE essentially involves computing the wavefunction under the action of the time evolution operator. Based on the energy–time uncertainty principle, the infinitesimal time step $dt$ can be approximated by $\Delta t$ without a significant loss of accuracy:
\begin{equation}\label{S1.4}
\Delta t=\frac{\hbar}{\Delta E}\approx dt,
\end{equation}
where $\Delta E$ is the maximum energy bandwidth of the wavefunction. By further applying a Taylor expansion and neglecting higher-order nonlinear terms, we obtain the final expression:
\begin{equation}\label{S1.5}
[1+\frac{i}{2\hbar}H(z,t)dt]\Psi(z,t+dt)=[1-\frac{i}{2\hbar}H(z,t)dt]\Psi(z,t).
\end{equation}
The core idea of the Crank-Nicolson method is to convert the Hamiltonian into a tridiagonal matrix. The Hamiltonian operator in a discrete grid is
\begin{equation}\label{S1.6}
H(z,t)\Psi(z,t)=-\frac{\hbar^2}{2m}\frac{\Psi(z+\Delta z,t)-2\Psi(z,t)+\Psi(z-\Delta z,t)}{\Delta z^2}-V(z,t)\Psi(z,t),
\end{equation}
where $\Delta z$ is the step of the spatial grid, and $V(z,t)$ is a potential. If the two spatial ends of the wavefunction are fixed, the corresponding matrix is
\begin{equation}\label{s1.7}
H(z,t)\Psi(z,t)=
\begin{bmatrix}
    \frac{\hbar^2}{m\Delta z^2}+V_1       &-\frac{\hbar^2}{2m\Delta z^2}&0&\cdots \\
    -\frac{\hbar^2}{2m\Delta z^2}       &\frac{\hbar^2}{m\Delta z^2}+V_2 & -\frac{\hbar^2}{2m\Delta z^2} &\ddots  \\ 0    & -\frac{\hbar^2}{2m\Delta z^2}   &  \frac{\hbar^2}{m\Delta z^2}+V_3 &\ddots\\ \vdots &\ddots&\ddots&\ddots
\end{bmatrix}
\begin{bmatrix}
\Psi_1\\ \Psi_2 \\ \Psi_3\\ \vdots
\end{bmatrix},
\end{equation}
where the subscript indicates the space indices. The evolved wavefunction is eventually obtained by
\begin{equation}\label{S1.8}
\Psi(z,t+dt)=[1+\frac{i}{2\hbar}H(z,t)dt]^{-1}[1-\frac{i}{2\hbar}H(z,t)dt]\Psi(z,t).
\end{equation}

In our numerical calculations, the maximum energy bandwidth is set to $\Delta E=50$ eV, which is approximately ten times higher than the cutoff energy of the tunnelling spectrum. Under this bandwidth, the temporal and spatial steps are $\Delta t=13$\,as and $\Delta z=27.6$\,pm, respectively. The ends of the spatial grid are set at $\pm 300$ nm, nearly 300 times the size of the nanojunction, effectively preventing unphysical reflections from the fixed boundaries.

\section{Effective Keldysh parameter for a two-colour field}\label{secKP}

The Keldysh parameter~\cite{Keldysh1965} is widely used to characterize ionization processes in strong-field physics. However, most theoretical frameworks primarily consider a single-frequency (monochromatic) driving field. In our work, the ratio $\eta$ between the second harmonic and fundamental field strengths is approximately
$\eta=\frac{F_2}{F_1}=32\ \%$, which is substantial and cannot be neglected.

\begin{figure}[htb!]
\centering
\includegraphics[width=0.5\linewidth]{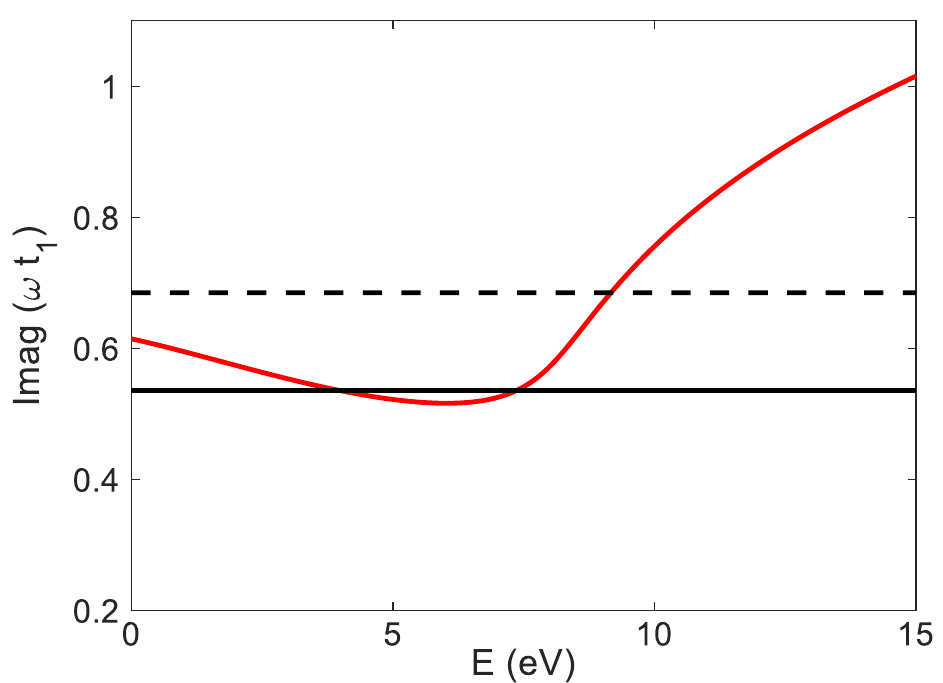}
\caption{Comparison between the Keldysh parameter and the numerically obtained saddle-point emission phase. The solid red curve represents the emission phase derived from the saddle-point equations as a function of the final transport energy. The solid black line corresponds to the modified Keldysh parameter, while the dashed black line shows the standard (unmodified) Keldysh parameter with $\eta=0$. The modified $\gamma$ provides a good description of the numerical solution up to the cutoff energy at 9.17\,eV. In these calculations, we use a field strength of $10\ \mathrm{V\,nm}^{-1}$ and a junction width of 1\,nm. All other parameters are consistent with the experimental conditions described in the main text.}
\label{FigSK}
\end{figure}

Since efficient current bursts are localized near the field crest~\cite{Ma2024, Ma2025}, we focus on the instantaneous field at this maximum. According to the saddle-point equation (Eq.~4 in the main text) and Eq.~\ref{ST5}, we derive the following approximation:
\begin{eqnarray}\label{SK1}
&&\tilde p+|e|\frac{F_1}{\omega}\big[\sin(\omega t_1)+\frac{\eta}{2}\sin(2\omega t_1)\big]\nonumber\\
&=&\tilde p+|e|\frac{F_1}{\omega}\sin(\omega t_1)\big[1+\eta\cos(\omega t_1)\big]\\
&=&i\sqrt{2m|E_0|_{\mathrm{eff}}}\nonumber,
\end{eqnarray}
where $|E_0|_{\mathrm{eff}}=|E_0|-\overline{|V_{\mathrm{imag}}|}$ denotes the effective initial energy. Comparing the imaginary components and performing a Taylor expansion of the cosine and hyperbolic cosine terms, we obtain the expression
\begin{equation}\label{SK2}
\sinh[\omega \Im(t_1)]=\frac{\omega\sqrt{2m|E_0|_{\mathrm{eff}}}}{|e|F_{1}(1+\eta)}=\gamma,
\end{equation}
where $\Im(t_1)$ is the imaginary component of the emission time. Thus, the effective Keldysh parameter for a two-colour field becomes:$\gamma=\frac{\omega\sqrt{2m|E_0|_{\mathrm{eff}}}}{|e|F_{1}(1+\eta)}$. Figure~\ref{FigSK} compares the modified and unmodified Keldysh parameters. The modified $\gamma$ provides a good approximation of the emission phase below the cutoff energy~\cite{Ma2025} (9.17 eV), whereas the unmodified parameter deviates from the numerical results.

\section{Dependence of two-colour current on junction width}
\label{sec:nearfield}

\begin{figure}[htb!]
    \centering
    \includegraphics[width=0.7\linewidth]{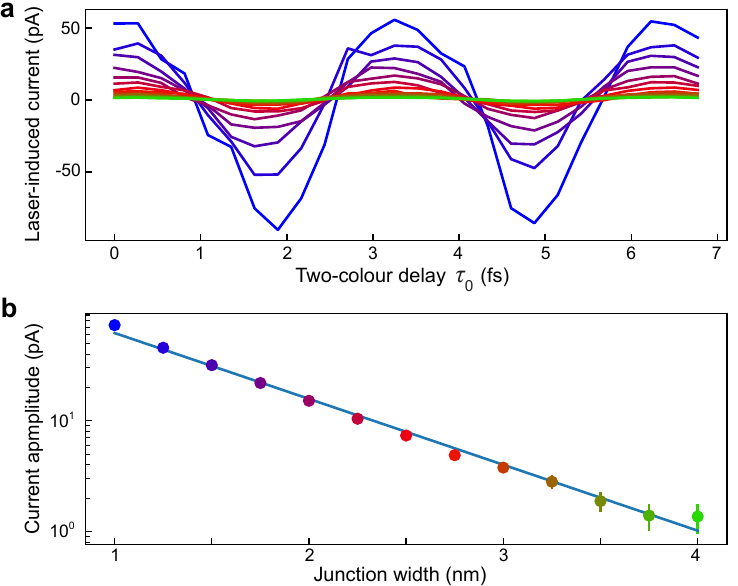}
    \caption{\REV{Experimental t}wo-colour delay scans for different junction widths. \textbf{a}, Laser-induced current (frozen tip, zero bias) for different junction widths as a function of the base two-colour delay $\tau_0$. \textbf{b}, Oscillation amplitude from \textbf{a} (coloured dots) as a function of the junction width. The blue line is an exponential fit. The colour of the dots is the same as the curves in a.}
    \label{FigDecay}
\end{figure}

In Fig.~\ref{FigDecay}a, we show \REV{experimental} two-colour delay scans as a function of the junction width for frozen tip and zero bias. Figure~\ref{FigDecay}b shows the two-colour modulation amplitude as a function of the junction width. We find an exponential decay of the amplitude (see fit curve). \REV{In Fig.~\ref{FigDecay}a, we observe a significant asymmetry in the current flow, leaning towards negative currents (sample to tip electron transfer). Such a trend is also visible in Fig.~2c and Fig.~4a of the main text, albeit much weaker. This asymmetry is explored with the TDDFT further below in Section~\ref{sec:TDDFT}. }

\section{Optical near-field enhancement simulations}\label{sec4}

\begin{figure}[htb!]
    \centering
    \includegraphics[width=1\linewidth]{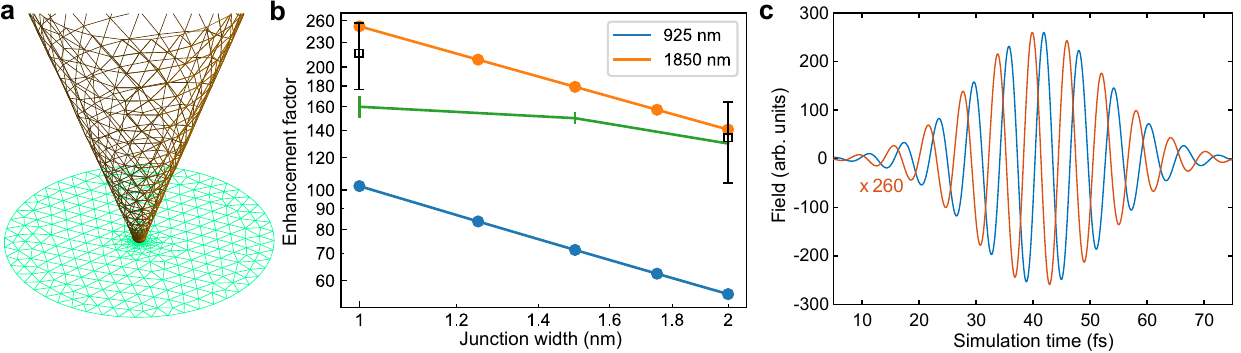}
    \caption{Maxwell simulations of the optical near-field. \textbf{a}, Mesh structure for the boundary element method (BEM) with $R=20$\,nm and junction width 1\,nm. \textbf{b}, BEM results for 1850\,nm (orange) and 925\,nm (blue) together with \REV{the estimated results derived from the experimental data (green: nTDSE; black: TDDFT; see main text).} \textbf{c}, Results of a finite-difference time-domain (FDTD) calculation for 1\,nm gap, 1850\,nm laser pulse. The orange curve is the field in free space, multiplied by a factor of 260. The blue curve represents the near-field in the junction.}
    \label{fig:Near}
\end{figure}

To calculate the expected field enhancement at the tip, we employed a boundary element method (BEM) simulation based on the BEMPP software package~\cite{Betcke2021} to numerically solve the Maxwell equations. In the calculation, the Pt:Ir tip was modeled as a half-sphere with radius $R$ connected to a cone with a half opening angle of 20 degrees. The length of the tip was chosen so that artificial effects, such as antenna resonances, are avoided. The gold sample was modeled as a flat disk with radius 400\,nm. The refinement of the mesh is focused around the apex of the tip, and approximately 2300 elements were used for the calculation with an area of 0.3 $\text{nm}^2$ around the tip apex. An example mesh is shown in Fig.~\ref{fig:Near}a. The incoming field is modeled as a plane wave at the central wavelength of the fundamental beam of 1850\,nm. The angle of incidence of the field corresponds to our experimental setup. The calculations are in fair agreement with the values \REV{extracted from the comparison of the experiment to the nTDSE model} (see Fig.~\ref{fig:Near}b). We also determined the near-field enhancement for the SH field at 925\,nm, which is roughly a factor of 2 weaker. \REV{We note that the local field near the tip is slightly stronger than the local field near the sample due to the sharpness of the tip. This is a major factor causing slight asymmetry in two-colour modulation experiments (see Fig.~4a in the main text and Section~\ref{sec:TDDFT}).}

\REV{We also estimated the field enhancement values when projecting the TDDFT calculation onto our measurement results in Fig.~3a in the main text. The TDDFT gives similar field enhancement values of $216 \pm 40$ (1\,nm junction) and $134 \pm 30$ (2\,nm junction), see the black open squares in Fig.~\ref{fig:Near}b.} 

In order to verify the veracity of the BEM simulation and the large field enhancement factors obtained, we performed a finite-difference time-domain (FDTD) simulation using the Lumerical software package for a 1850\,nm pulse with a duration of 20\,fs and a junction width of 1\,nm. We find a field enhancement factor of 260, in good agreement with the BEM simulation. The time domain field (see Fig.~\ref{fig:Near}c) also reveals the absence of plasmon ringing.

\section{Robustness of current burst durations}

\begin{figure}[htb!]
    \centering
    \includegraphics[width=0.85\linewidth]{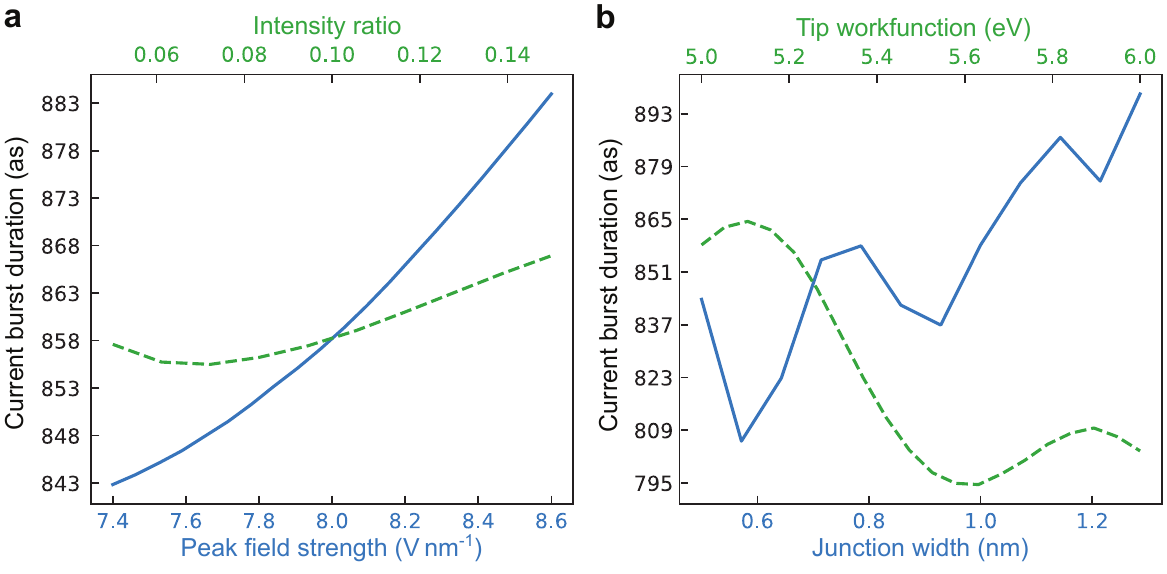}
    \caption{Robustness of attosecond current burst durations.  \textbf{a}, FWHM duration of the main current burst at the vacuum-sample boundary as a function of peak field strength (solid blue curve) and intensity ratio between SH and fundamental (green dashed curve). \textbf{b}, The same, but as a function of junction width (solid blue curve) and the workfunction of the nanotip (dashed green curve).}
    \label{fig:Variation}
\end{figure}

\REV{In order to assess the robustness of our results, we show below the comparison between the nTDSE and TDDFT results (see Fig.~\ref{fig:Duration}) which shows quantitative agreement between one-electron model and many-body theories. We also} perform nTDSE calculations and vary parameters around a specific set of experimental parameters (peak field strength 8\,V\,nm$^{-1}$ \REV{of the total field}, 10\% intensity ratio, junction width 1\,nm, workfunction of nanotip 5.1\,eV). Figure~\ref{fig:Variation} shows the FWHM current burst duration as a function of four crucial parameters. Remarkable is the behaviour with peak field strength. One would assume that higher field strength and lower $\gamma$ would further decrease the current burst duration, but here the opposite is the case (Fig.~\ref{fig:Variation}a, solid blue curve). The underlying reason is that the nonlinearity of tunnelling decreases with increasing field strength. The dependence of the burst duration on the intensity ratio of SH and fundamental is extremely weak  (Fig.~\ref{fig:Variation}a, dashed green curve). Increasing the workfunction of the nanotip leads to an overall decrease in burst duration due to an increase in nonlinearity (Fig.~\ref{fig:Variation}b, solid blue curve). The junction width dictates the travel time of the attosecond electron wavepacket; its increase results in an overall increase in the dispersion and hence the duration (Fig.~\ref{fig:Variation}b, dashed green curve). We conclude that the sub-femtosecond confinement of the waveform-controlled sub-cycle current bursts remains robust against reasonable variations of experimental parameters.

\REV{The variation of the current burst duration obtained from the nTDSE calculation is mostly affected by uncertainty in the junction width $d$ (see Fig.~\ref{fig:Variation}b) and the field enhancement (see Fig.~\ref{fig:Variation}a and Fig.~\ref{fig:Near}b). Assuming an uncertainty of $d$ of $\pm 0.3$\,nm and considering the error in determining the field enhancement, we can estimate a model-based uncertainty of the current burst duration of $\pm 90$\,as.}

\section{Dependence of current directionality on second-harmonic intensity}

\begin{figure}[htb!]
    \centering
    \includegraphics[width=0.55\linewidth]{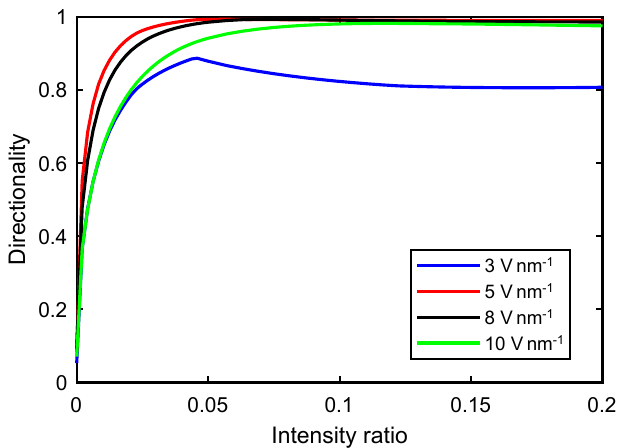}
    \caption{Directionality as a function of the ratio of the SH intensity to the fundamental intensity for different peak fields (see legend).}
    \label{fig:Direct}
\end{figure}

Figure~\ref{fig:Direct} shows the dependence of the induced symmetry breaking in the current flow on the intensity ratio of the SH field to the fundamental field, quantified by the directionality $\Delta = |(J_{+} - J_{-})/(J_{+} + J_{-})|$. Here, $J_{+}$ and $J_{-}$ are the integrated currents in the positive and negative directions, respectively. Starting at low intensity ratios, $\Delta$ increases strongly and reaches a saturation regime already around an intensity ratio of 0.06 for peak fields $5\,\mathrm{V\,nm}^{-1}$ and $8\,\mathrm{V\,nm}^{-1}$. 

\section{Carrier-envelope phase effects in two-colour STM experiments}

\begin{figure}[htb!]
    \centering
    \includegraphics[width=0.95\linewidth]{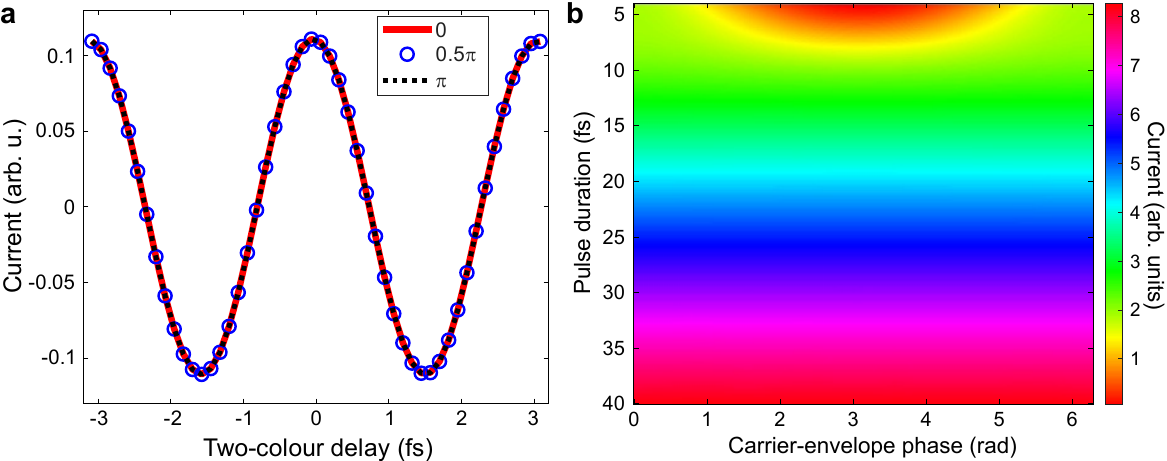}
    \caption{Carrier-envelope phase (CEP) effects calculated with the nTDSE model. \textbf{a}, Two-colour delay scan of the current for three CEP values (see legend). \textbf{b}, CEP dependence of the current for zero two-colour delay and varying pulse duration.}
    \label{fig:CEPtdse}
\end{figure}

In our experiment, the two-colour field breaks the symmetry and induces net current flow with controllable magnitude and direction. The carrier-envelope phase (CEP) of our laser pulses is not stabilized. However, CEP effects may induce an additional modulation if the pulses are sufficiently short, typically in the few-cycle regime. In order to assess the magnitude of these effects, we performed TDSE simulations comparing the two-colour delay-dependent current for CEP values of 0, 0.5$\pi$\,rad, and $\pi$\,rad, respectively. No significant differences are observed in the calculated current shown in Fig.~\ref{fig:CEP}a as the CEP effect leads to a relative modulation of much less than 1\,\%. Moreover, varying the fundamental pulse duration from 40\,fs to 4\,fs (less than one optical cycle) and the CEP from 0 to 2$\pi$\,rad at a fixed two-colour delay of 0 shows a clear CEP effect can be observed only below 10\,fs (see Fig.~\ref{fig:CEPtdse}b). Therefore, for the pulse durations used in our experiment, CEP effects are negligible.

\section{Influence of second-harmonic chirp on the two-colour modulation}

Our measurements were performed around the maximum of the overlap between fundamental and second harmonic, which is near the temporal center of the pulses. As shown in Fig.~2c in the main text, no visible variation in the current amplitude is observed within the measured delay window. For a large delay scan (Fig.~\ref{figLong}), higher-order chirp-induced tails become visible beyond two-colour delays of 25\,fs. Such chirp modifies the instantaneous frequency and may therefore influence frequency-dependent parameters. In our experiment, the second harmonic is introduced primarily to break the field symmetry. A frequency shift due to chirp does not significantly alter the generation of individual current bursts, but it can slightly modify the period of the net current reversal. Parameters that explicitly depend on the laser frequency, such as the Keldysh parameter, may therefore exhibit deviations in the theoretical analysis.
Importantly, the reconstruction of the current from lock-in measurements is sensitive to the instantaneous field frequency. Nevertheless, Fig.~2c in the main text shows that the reconstructed laser-induced current agrees well with the directly measured current within the experimentally probed delay range. This consistency indicates that chirp-induced deviations are negligible in the domain of our measurements.

\begin{figure}[htb!]
    \centering
    \includegraphics[width=0.95\linewidth]{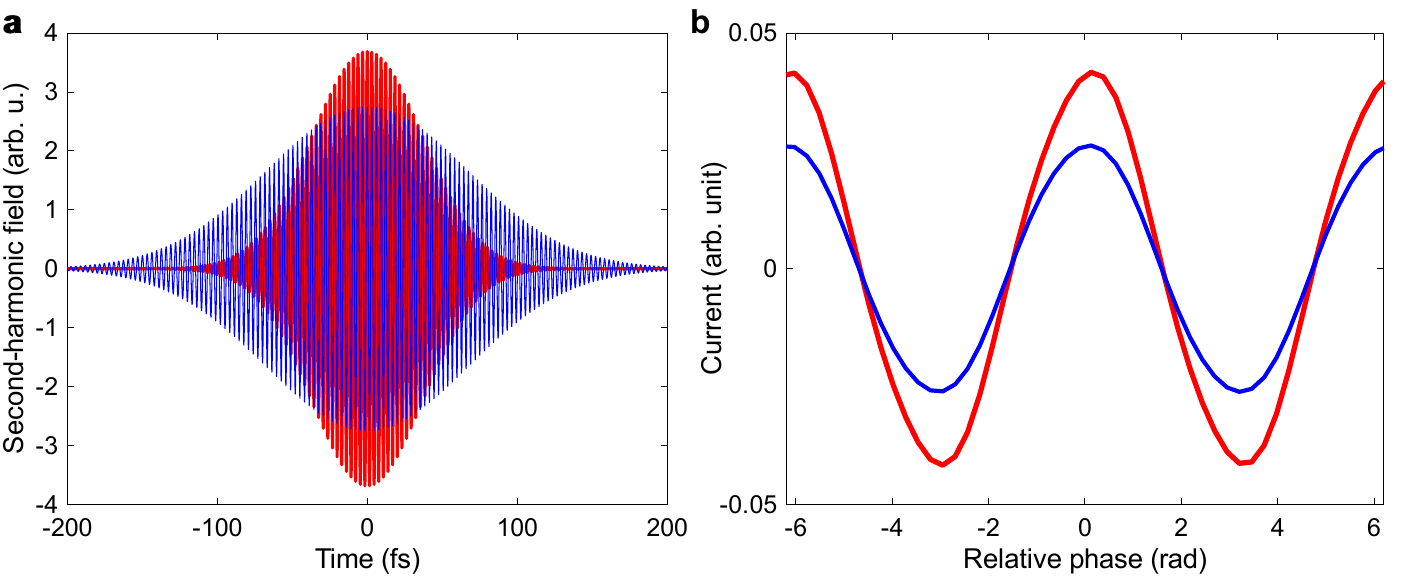}
    \caption{Effect of second-harmonic chirp calculated with the nTDSE model. \textbf{a}, Unchirped (blue) and chirped second-harmonic field (red). \textbf{b}, Two-colour delay scan of the current for unchirped (blue) and chirped (red) second harmonic. The peak field strength in this calculation is 8\,V\,nm$^{-1}$.}
    \label{fig:chirp}
\end{figure}

Another effect of chirp is a decrease of the amplitude of the current oscillation in two-colour delay scans.
In Fig.~\ref{fig:chirp}, we show TDSE calculations after introducing a second-order chirp into a Fourier-limited second-harmonic pulse, increasing its duration from 62\,fs to 112\,fs. The second harmonic field is given by 

\begin{equation*}
{\cal E}_2(t) = (1 + C^2)^{-1/4} F_2 \exp \left[-\frac{2\ln(2)}{\tau_2^2 (1+C^2)}t^2\right] \cos{\left( \omega t + \frac{C}{\tau_2^2  (1+C^2)} t^2\right)}.
\end{equation*}

Here, we use $\tau_2 = 62$\,fs and $C = 1.5$ ($C = 0$) for the chirped (unchirped) pulse.

Due to a drop in total peak field strength, the amplitude of the two-colour oscillations also drops. In the time-domain, such a drop can cause small variations of the attosecond current burst duration on the order of $\pm 10$\,as (see dashed green curve in Fig.~\ref{fig:Variation}a).

\section{Robustness of the experiment with respect to instabilities}

STM instabilities can affect the measurements through several avenues and require mitigation strategies, which we will discuss in the following:

\begin{itemize}

\item \textit{STM drift of the lateral position (x / y).} We routinely observe drifts on the order of 2nm/min in our system. The strategy for enabling long-time measurements using the tip freezing approach is to first perform a large area scan and then select a flat grain of the gold sample (sizes of up to 200 x 200 nm$^2$) which produces a homogeneous laser-induced lock-in signal. As long as we stay on the grain, our laser-induced measurements are not affected. If the tip slides off the grain, strong changes to the signal and the attosecond current dynamics can ensue due to (i) the strong sensitivity of the laser-induced current to grain boundaries (see Fig. 1c in the main text), (ii) sensitivity to local work function changes due to local variations in crystallographic orientation, and (iii) concomitant changes in the optical near-field enhancement factor.

\item \textit{Mechanical vibrations.} We measured the vibration level on our optical table next to the STM setup to be between VC-D and VC-E ($\sim$4 $\mu$m/s RMS vibration velocity). Both lock-in approach (uses the STM feedback loop continuously) and tip freezing (uses the STM feedback loop intermittently to bring the tip back into position relative to the sample) are able to provide us with sufficient stability to record data in a long-term measurement.

\item \textit{Feedback bandwidth. }We chose an STM feedback-loop bandwidth of 400 Hz (constant current mode) and a lock-in modulation frequency which is nearly a magnitude higher. We do not observe any cross-talk between feedback loop and laser-induced lock-in signal.

\item \textit{Instability of the two-colour delay.} We actively monitor the stability of the two-colour delay by splitting off a part of the two-colour light and sending it to another setup where the fundamental pulse drives another second harmonic process. The latter is interfered with the second harmonic from the two-colour pulse with the help of a polarizer. We monitor the resulting spectral interference fringes.

\end{itemize}

In order to estimate an upper bound on attosecond sensitivity of our experimental system under the above conditions, we analyze the lock-in phase map of Fig.~\ref{figPhase}, which includes a grain boundary. We find that the lock-in phase RMS is 5 degrees, corresponding to 45 as RMS in the time domain.

Future atomic-scale imaging of small objects, such as molecules, will require much lower STM drift. A cryogenic ultrahigh-vacuum STM system will readily provide low drift.

\section{Possible experimental avenues for accessing STM electron energy spectra}

As stated in the main text, STM detection does not provide direct energy resolution. Therefore, the energy-related interpretations presented in our work need to rely strongly on comparison between experimental current measurements and theoretical modelling. Further experimental verification of the predicted energy distribution of transmitted currents will require introducing additional control parameters. One possible approach is a bias-dependent measurement. Applying a static bias across the tip–sample junction modifies the tunnelling barrier and photoemission rate~\cite{Schroder2020}. A map of the laser-induced conductance $dI/dV$ as a function of the two-colour delay could help to retrieve the emission process, as the added static field lowers the effective barrier and enhances electron emission. Another promising route would be combining a quasi-static THz field with a near-infrared driving field to modulate the barrier shape in a controlled manner, enabling further investigation of the light-electron interactions~\cite{Muller2020}. Alternatively, it is conceivable that the detection of light emission from the STM junction may provide information about the electron energy distribution~\cite{Siday2024}.

\section{Time-dependent density functional theory calculations of electron transport}
\label{sec:TDDFT}

Atomic units (a.u.) are used in this section unless otherwise stated. 

\subsection{Model and formalism}

Our many-body calculations of electron transfer in the STM junction, coherently 
controlled by two-colour laser pulses, are based on time-dependent density 
functional theory (TDDFT). The method has been employed in a number of studies of 
optically induced electron transfer in general, and photon-assisted tunnelling in 
particular, in both wide and narrow vacuum gaps between metals~\cite{Marinica12,
TeperikCylDimer,Ludwig2019,HotElectrons_in_Gaps,PAT_PRA_2026}. Its detailed 
description can therefore be found in previous works \cite{TeperikCylDimer,
Ludwig2019,HotElectrons_in_Gaps,PAT_PRA_2026} (see the main text of these 
references and their Supplementary Information).

Electron transport across the STM junction between metals, triggered 
by a strong optical field, is to a large extent determined by the potential 
landscape at the metal/vacuum interface probed by delocalized valence electrons, 
and by the dynamics of electron currents in the (vacuum) gap between metals. 
Therefore, free-electron modelling of the metal band structure has proven to be 
very efficient for the description of the experiments on strong-field photoemission and
transport~\cite{Lemell_CEP_2003,PDombi2004,Apolonski04,Kruger2012,Wachter2012,
Ludwig2019,PhysRevBbias,Light_STM_Kumagai23,RevModPhys.92.025003}
Within the free-electron (jellium) model, the ions at the lattice sites are 
represented by a uniform positive background charge density 
$n_+=\left[\frac{4 \pi}{3} r_s^3\right]^{-1}$ confined within the geometrical 
surface of the material. The Wigner-Seitz radius $r_s=3.02$~a$_0$, typical 
for silver and gold, is used in our work (a$_0$=1~a.u. $\approx 0.529$~\AA~is the Bohr radius).

The advantage of the free-electron model is that one can address systems of 
relatively large size, relevant for the experiments, which are out of reach 
for quantum approaches based on the \textit{ab initio} modeling of materials  
\cite{STM_CEP_Mukamel,PhysRevB.105.085416,Chen2018}. 
That said, even jellium modeling of metals does not allow one to reproduce 
a three-dimensional (3D) junction formed by an STM tip of 10-20~nm radius 
and the flat metal surface. To mimic the geometry effects in the actual system,
we therefore perform a two-dimensional (2D)  
calculations, considering optically induced electron currents across a  
vacuum gap of size $d$  in a dimer system formed by two parallel cylindrical 
nanowires with radii $R_1 = 3.3$~nm (representing the STM tip), 
and $R_2 = 7$~nm (representing the sample surface), as sketched in the 
inset of Fig.~\ref{fig:y-dependent-J}. The $x$-axis is chosen along the 
dimer axis connecting the centers of the nanowires, and the $z$-axis is along 
the nanowires. Since the system has translational invariance along the 
$z$-axis, all quantities characterizing the electron dynamics are obtained  
per unit ($1$~a$_0$) length of the dimer.

Within the present implementation of TDDFT, the time-dependent 2D electron 
density $n(\mathbf{r},t)$, is represented as a sum over the occupied 
Kohn--Sham (KS) orbitals $\Psi_k(\mathbf{r},t)$ 
\begin{align}\label{eq:density}
n(\mathbf{r},t) &= \sum_{k \subset occ} 
\chi_k \left|\Psi_k(\mathbf{r},t)\right|^2, 
\end{align}
where $\mathbf{r}=(x,y)$.
The statistical factor $\chi_k$ accounts for the $\pm \frac{1}{2}$ 
electron spin degeneracy, and the degeneracy associated with motion 
along $z$-axis.

The KS orbitals evolve in time according to the time-dependent KS 
equations \cite{MarquesGross}
\begin{align}\label{eq:TDDFT_KS}
i \partial_t \Psi_k(\mathbf{r},t) = \left( \hat{T} + V(\mathbf{r},t) \right) 
 \Psi_k(\mathbf{r},t).
\end{align}
Here, $\hat{T} = -\frac{1}{2} \left(\frac{d^2}{dx^2} + \frac{d^2}{dy^2}  \right)$
is the kinetic energy operator, and the effective one-electron potential 
$V(\mathbf{r},t)$ consists of four contributions (in order of 
appearance on the right-hand side of Eq.~\ref{eq:TDDFT_Potentials}): the Hartree potential, 
the exchange--correlation potential, the stabilization potential, and the potential of the 
incident optical pulse.
\begin{align}\label{eq:TDDFT_Potentials}
V(\mathbf{r},t) = \int \frac{n(\mathbf{r}',t)}{|\mathbf{r}-\mathbf{r}' |}  \ d^3 \mathbf{r}' +V_{\rm{XC}}(\mathbf{r},t)+v_{\rm{s}}(\mathbf{r})+V_{\rm{opt}}(x,t)
\end{align}
The Hartree potential and the exchange--correlation potential, 
$V_{\rm{XC}}(\mathbf{r},t)$, depend on time through the time-dependent 
electron density. For $V_{\rm{XC}}(\mathbf{r},t)$, we use the adiabatic 
local density approximation (ALDA) \cite{MarquesGross} with the 
exchange--correlation kernel of Gunnarsson and Lundqvist \cite{gunnarsson1976exchange}. 
The choice of the stabilization potential, 
$v_{\rm{s}}(\mathbf{r})=v^{(1)}_{\rm{s}}+v^{(2)}_{\rm{s}}$ within the stabilized 
jellium model \cite{perdew1990stabilized} of metals allows one to set 
the work functions $\Phi_1$ of cylinder 1, and $\Phi_2$ of cylinder 2 independently. 
The stabilization potential for each individual cylinder is given by
\begin{align}\label{eq:TDDFT_Stab}
v^{(j)}_{\rm{s}}(r_j) =  - A_j/(1+e^{r_j-R_j}),~~~~j=1,2,
\end{align}
where $r_j$ is the radial coordinate with respect to the center of cylinder $j$, 
and $A_j$ is the amplitude. Unless otherwise stated, the results reported 
in the main text and in this SI are obtained with $\Phi_1=\Phi_2=5.1$~eV, in line 
with the one-dimensional (1D) nTDSE model. Test calculations with $\Phi_1=\Phi_2=4.5$~eV, 
demonstrate the quantitative robustness of the characteristic times 
reported here for the dynamics of electron currents in the junction. 
To assess the possible role of differences between the tip and surface work 
functions, we also considered the situation with $\Phi_1=5.5$~eV for the Pt:Ir tip, 
and $\Phi_2=5.1$~eV for the gold surface (see below).

The potential of the optical pulse in the length gauge is 
$V_{\rm{opt}}(x,t) = x~\mathcal{E}(t)$, where $\mathcal{E}(t)$ is the 
electric field of the incident fundamental wave and its second harmonic. 
The polarization of the incident optical pulse is such that the electric 
field vector is aligned with the dimer axis ($x$-axis).
The experimental pulse durations of 35~fs (fundamental) and 62~fs (second 
harmonic) fields would lead to prohibitively long computation times. 
We therefore model the free-space electric field as 
\begin{equation}
\label{Eq:ElectricField}
\mathcal{E}(t) = -F \ e^{-2\ln(2)\left(\frac{t}{\tau}\right)^2} \
 \left[ \cos\left(\omega t  \right) 
 + A_{2\omega} \cos\left(2\omega t - \phi \right) \right],
\end{equation}
where $\phi$ is the relative phase, and, consistently 
with experimental conditions, $A_{2\omega}=1/\sqrt{10}$. Performing the calculations with the 
fundamental field only, we have explicitly checked that using a pulse 
duration $\tau=15.4$~fs is sufficient to ensure that the optical pulse is 
long enough  so that the optically induced transport is independent of 
the carrier-envelope phase (CEP) \cite{Ma2025,PAT_PRA_2026}. 
A common exponential factor in Eq.~\ref{Eq:ElectricField} allows one to 
avoid the effect of the shift of the envelope of the second harmonic with 
respect to the fundamental pulse. This effect is absent in the experiment due to  
significantly longer pulse durations.

The KS orbitals $\Psi_k(\mathbf{r},t)$ are represented on an equidistant 
mesh in the $(x,y)$ coordinates (mesh step $0.8$~a$_0$), and Eqs.~\ref{eq:TDDFT_KS} 
are solved using the Fourier-grid technique \cite{kosloff1983fourier} and 
the short-time split-operator propagation \cite{Leforestier1991} (time step 
$\Delta t=0.125$~a.u.). 
The initial conditions $\Psi_k(\mathbf{r},t=T)$ are given by the 
occupied KS orbitals of the ground-state system. The starting time for the time propagation,  
$T$, is chosen such that $E(T)\approx 0$ (typically $T=-4 \tau/\sqrt{2\ln(2)}$).

\subsection{Laser-induced electron transport }

Along with the time-dependent electron density $n(\mathbf{r},t)$, 
we also calculate the time-dependent electron current density $\mathbf{j}(\mathbf{r},t)$ 
\begin{align}\label{eq:CurrentDens_J}
\mathbf{j}(\mathbf{r},t) &=\sum_{k \subset occ} 
\chi_k \text{Im}\left[\Psi^*_k(\mathbf{r},t)\mathbf{\nabla}\Psi_k(\mathbf{r},t)\right],  
\end{align}
where $Z^*$ stands for the complex conjugate of $Z$, 
and $\text{Im}\left[Z\right]$ stands for the imaginary part of $Z$. 
Finally, the net electron transfer $\mathcal{N}$ between the cylinders per 
optical pulse and per $1$~a$_0$ length of the dimer is given by 
\begin{equation}\label{eq:NetTransfer}
  \mathcal{N}=\int \,dt \, \underbrace{\int \,dy ~ 
  \hat{e}_x \cdot \mathbf{j}(x=0,y,t)}_{\mathcal{I}_e(x=0,t)}, 
\end{equation}
where $\hat{e}_x$ is the unit length vector in the positive direction of the $x$-axis, 
and $x=0$ corresponds to the middle of the junction. The inner integral is the 
total electron current $\mathcal{I}_e(x=0,t)$ through the middle of the junction. 
Given the laser pulse repetition rate $f$, the laser induced current 
in the junction is obtained from $\mathcal{I}_{\rm{optical}}=-\mathcal{N} f$, 
where the minus sign accounts for the negative sign of the electron charge.
We observe that with the choice of the optical pulse in Eq.~\ref{Eq:ElectricField}, 
$\mathcal{N}$ is positive and $\mathcal{I}_{\rm{optical}}$ is negative 
(electrons are transferred in the positive direction of the $x$-axis from the tip  
to the surface) for the zero relative phase between $\omega$ and $2 \omega$ fields, 
i.e. for $\phi=0$. The experimental definition implies the shift of $\phi$ by $\pi$.

\subsection{Relating the 2D model with the 3D experimental system and estimating the effective transport area}

\begin{figure}[htb!]
\centering
\includegraphics[width=0.8\linewidth]{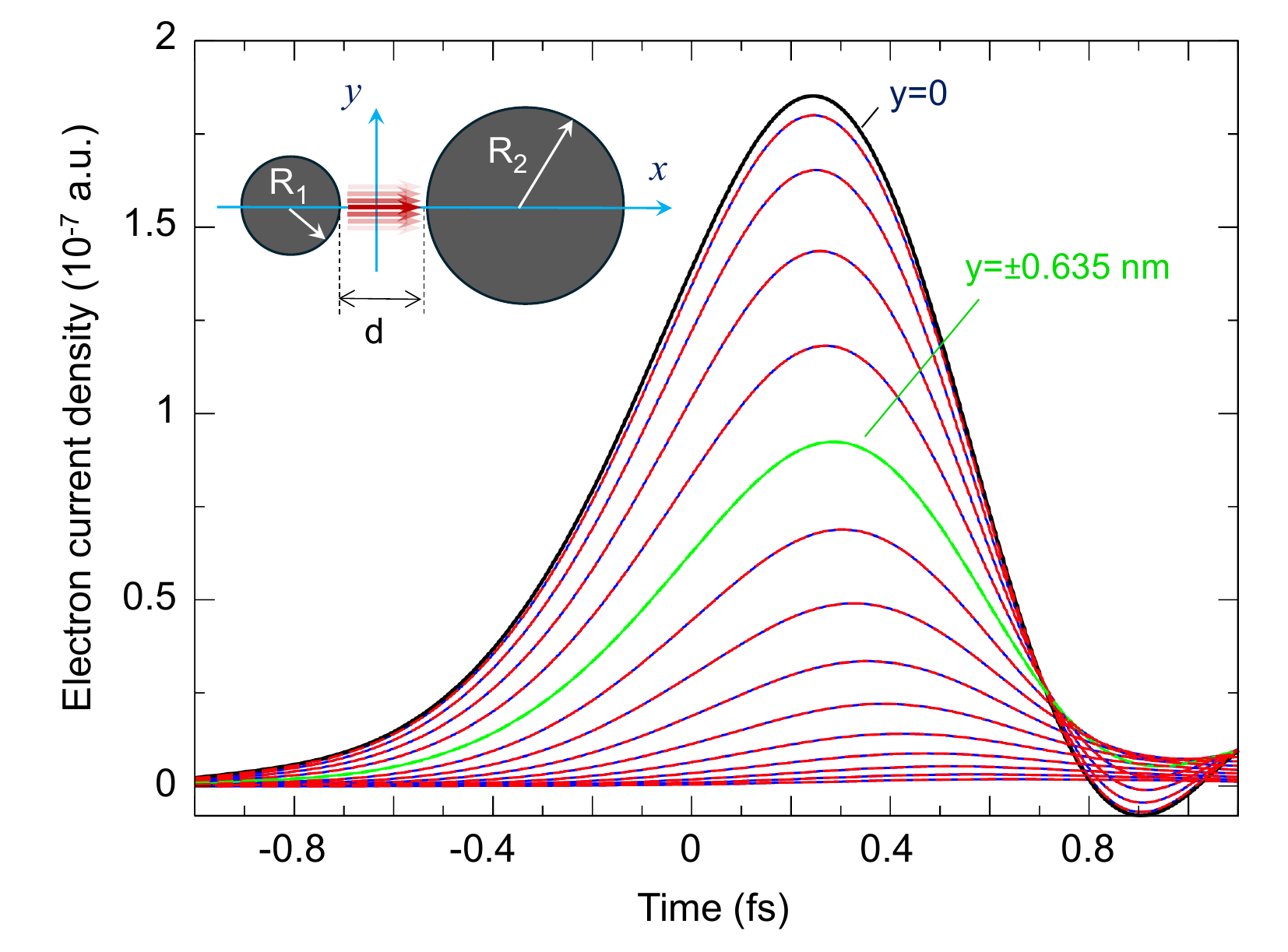}
\caption{
The $x$-component of the electron current density $j_x(x=0,y,t)$ induced by 
a $\phi = 0$, two-colour optical pulse in the middle of a $d=1$~nm wide junction 
between cylindrical nanowires. The time interval corresponds to the main crest 
of the field in the gap, with field value $\mathcal{E}_g = 9.55\,\mathrm{V\,nm}^{-1}$ reached 
in the middle of the junction. 
Results are shown as a function of time for different values of $y$-coordinate. 
Black line: electron current density along the dimer axis $j_x(x=0,y=0,t)$. 
Dashed red (solid blue) lines: results obtained for positive 
(negative) $y$-values spaced by 2.4~a$_0$. Green line: $j_x(x=0,y_{\frac{1}{2}},t)$ 
reaching half of the maximum value at the dimer axis, and calculated for 
$y_{\frac{1}{2}}=\pm 12$~a$_0$ ($\pm 0.635$~nm). 
The inset sketches the cut of the nanowire dimer by the $(x,y)$- plane, which 
allows to introduce the parameters of the system geometry: nanowire radii $R_1$ 
and $R_2$, and the size of the gap $d$.  
The nanowires are infinite along $z$-axis. Red arrows indicate electron transfer. 
}
\label{fig:y-dependent-J}
\end{figure}

In order to quantitatively compare laser induced currents calculated for 
the model 2D system and experimentally measured in STM junctions with 
a 3D geometry, one has to determine the effective transport area of the 
model system and that of the actual device. 
To this end, we analyze in Fig.~\ref{fig:y-dependent-J} the $y$-coordinate 
dependence of the optically induced electron current density 
$j_x(x=0,y,t)=\hat{e}_x \cdot \mathbf{j}(x=0,y,t)$. Results for different 
$y$ are shown as a function of time in a window spanning the region close 
to the main crest of the optical field, where the induced electron current density 
is strongest. 
Since the tunnelling barrier of the junction is smallest close to the 
dimer axis, $j_x(x=0,y,t)$ maximizes for $y=0$, and it rapidly decreases 
with increasing $y$. The effective FWHM of the calculated $y$-distribution 
of $j_x(x=0,y,t)$ is $L_y=24$~a$_0$  (1.27~nm). Considering that this 
result merely reflects the properties of the tunnelling barrier and 
therefore can be used for the different field strengths, the average laser-induced 
current density can be expressed as  
$\mathcal{I}_{\rm{optical}}/\mathcal{S}_{\rm{TDDFT}}$.  
Here $\mathcal{S}_{\rm{TDDFT}}=24$~a$_0^2$ is an effective transport area 
of our model system. Indeed, $\mathcal{I}_{\rm{optical}}$ is calculated 
with TDDFT per $1$~a$_0$ length of the dimer.  
Given an estimate for an effective transport area in the experimental device, 
$\mathcal{S}_{\rm{exp}}$, the theoretical prediction to be compared with the
experiment is given by $\mathcal{I}_{\rm{optical}}\ 
\frac{\mathcal{S}_{\rm{exp}}}{\mathcal{S}_{\rm{TDDFT}}}$. This comparison of TDDFT theory and experiment, also taking into account the difference in pulse durations, yields an effective transport area of $\sim$8\,nm$^{2}$.

It is worth noting that the dipolar plasmon of the free-electron 
metal nanowire dimer is off-resonance with the incident optical 
pulse. This is consistent with classical electromagnetic simulations for 
the present system and it allows one to avoid the plasmon ringing. 
The field in the middle of the junction in this situation is given by  
\begin{equation}\label{eq:Egap}
  \mathcal{E}_{\rm{g}}(t) = \mathcal{R} \mathcal{E}(t), 
\end{equation}
where $\mathcal{R}$ is the field enhancement factor. From the TDDFT 
calculations, we obtain $\mathcal{R}=3.85$ for the 1~nm junction, 
and $\mathcal{R}=2.74$ for the 2~nm junction. Primarily because 
of the difference between the model geometry and the experimental one, 
these values of $\mathcal{R}$ are about two orders of magnitude 
smaller than those obtained from classical electromagnetic simulations. 
In this respect, comparison between the calculated and measured laser-induced 
current allows an independent estimate of the amplitude of the 
field in the gap and thus of the field enhancement in the experimental
system.

\subsection{Attosecond dynamics of currents in the TDDFT}

\begin{figure}[htb!]
\centering
\includegraphics[width=0.75\linewidth]{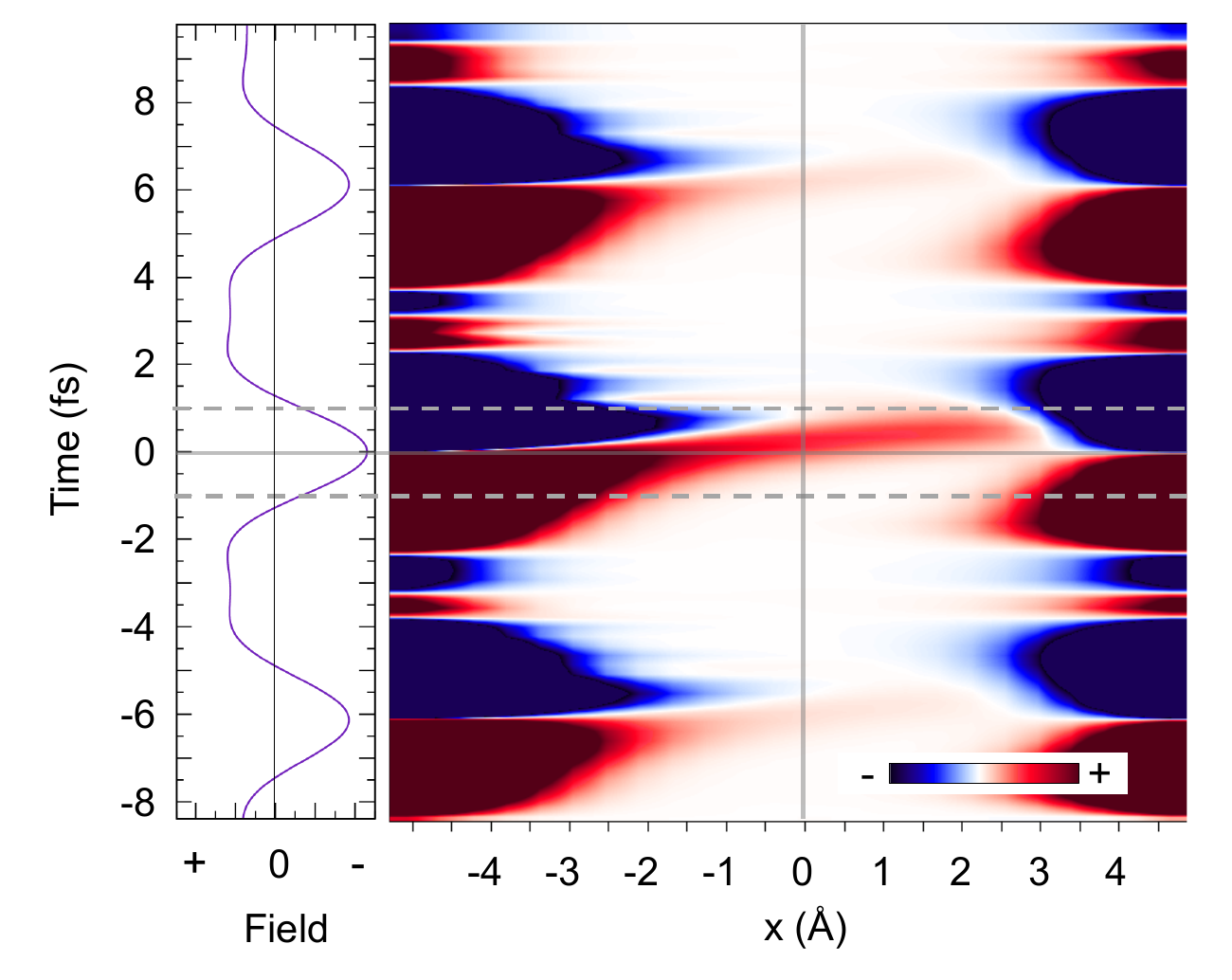}
\caption{Dynamics of electron currents induced by a $\phi = 0$ two-colour 
optical pulse in a $d=1$~nm junction between cylindrical nanowires. The 
colour map displays the $x$-component of the electron current density 
calculated with TDDFT along the dimer axis, 
$j_x(x,y=0,t)=\hat{e_x}~ \mathbf{j}(x,y=0,t)$.   
The TDDFT results are presented as a function of $x$-coordinate 
(horizontal axis) and time (vertical axis). The colour code 
is explained in the inset. At the main crest of the optical field 
($t=0$), indicated with a horizontal grey line, the optical field in 
the junction reaches $\mathcal{E}_g = 10.3\,\mathrm{V\,nm}^{-1}$ and leads to 
the strongest burst of electrons injected in the gap. 
The line plot at the left shows the time evolution of the electric 
field in the middle of the gap, $E_{\rm{gap}}(t)$. 
Horizontal dashed grey lines indicate the reference time instants ($t=\pm 1$~fs).
Vertical grey line marks the middle of the gap.
}
\label{fig:Interpolated}
\end{figure}

In Fig.~\ref{fig:Interpolated}, we show the colour map of the time evolution 
of the electron current density $j_x(x,y=0,t)=\hat{e_x}~ \mathbf{j}(x,y=0,t)$ 
calculated with TDDFT along the axis of the dimer with 1~nm gap. 
The electron dynamics in the system is triggered by the incident two-colour 
pulse with relative phase $\phi = 0$. 
As follows from the TDDFT results, at the crests of the optical field with 
negative polarity, electron bursts detach from the surface of the left 
cylinder and cross the gap in the positive direction of the $x$-axis on the 
time scale below 1~fs. The most intense electron burst is injected into the 
gap close to $t=0$, when the  maximum optical field $\mathcal{E}_g = 10.3\,\mathrm{V\,nm}^{-1}$ 
is reached in the junction. Because of the strongly nonlinear character of the 
process, electron emission produced by the field crests with positive 
polarity is invisible on the scale of the figure.

Interestingly, along with the optically induced inter-nanowire electron 
current density associated with net electron transport across the gap, 
the intense optical field induces strong polarization currents. The latter 
reflect the intra-nanowire process in which electrons in the individual 
nanowires flow towards their surfaces to screen external electric field. 
Because of electron wave function spill-out, the 
time-oscillations of the screening (polarization) charge at metal surfaces 
across the gap appear as electron currents inside the gap region, 
approximately for $|x|>2.5~$~\AA. As an important observation also 
reported in a recent paper \cite{Siday2024}, the transport and 
polarization currents are shifted by half of the optical period. Indeed, 
the transport current is strongest with electron burst injection into 
the gap when the field in the gap reaches its maximum (and consequently 
when screening charges are also maximal). At variance, the polarization 
current, associated with time derivative of the screening charges 
is strongest at zero crossing of the optical field.

\begin{figure}[htb!]
\centering
\includegraphics[width=0.7\linewidth]{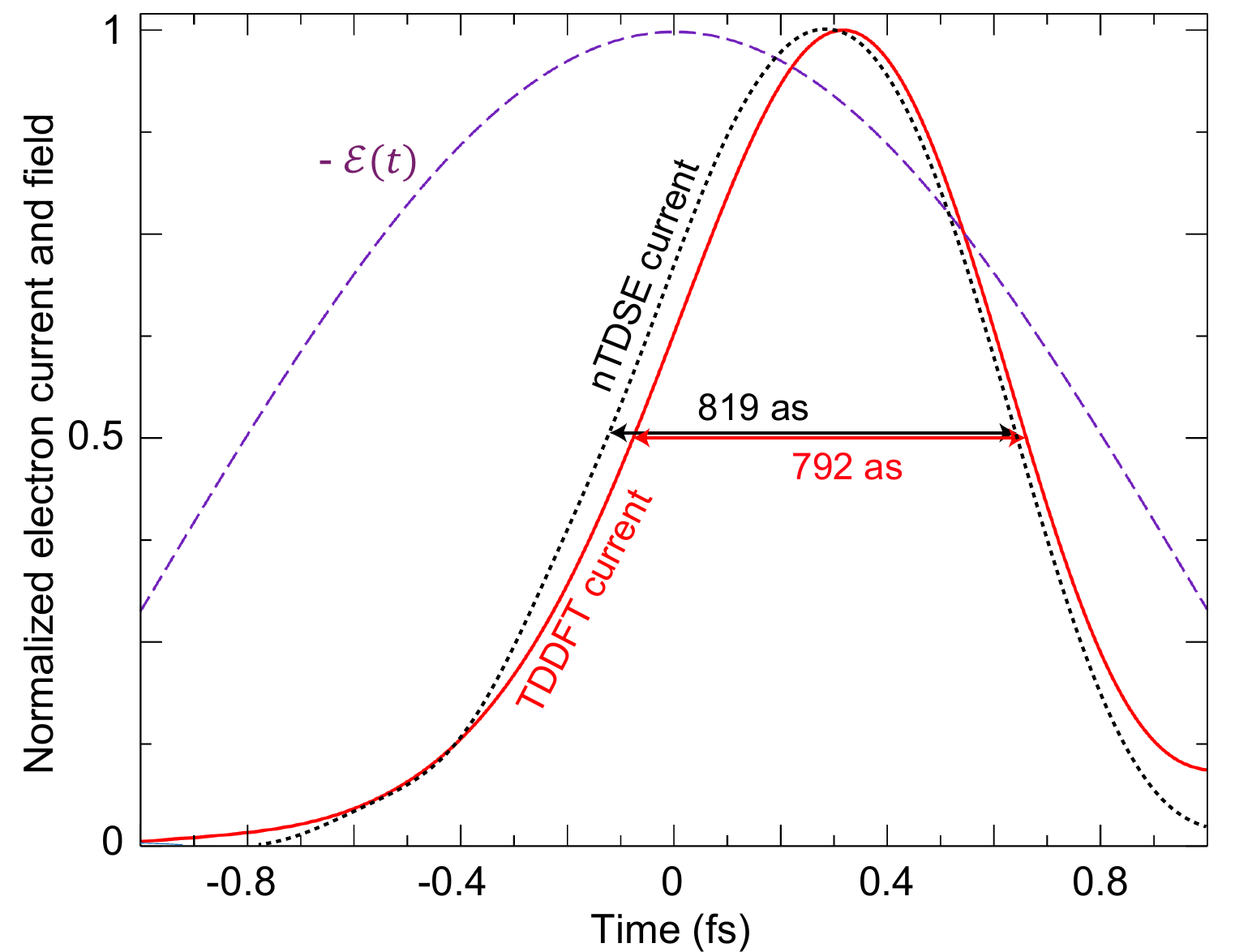}
\caption{Electron current $\mathcal{I}_e(x=0,t)$ in the centre of the gap between the nanowires, 
induced by a $\phi = 0$ two-colour optical pulse. The maximum field 
reached in the 1~nm gap is $10.27\,\mathrm{V\,nm}^{-1}$. Red solid curve: 
$\mathcal{I}_e(x=0,t)$ calculated with TDDFT (red solid curve); black dotted curve: the current 
profile obtained from the nTDSE calculations; dashed violet 
curve: the optical field $\mathcal{E}(t)$ multiplied by $-1$ for the sake of presentation. Results are shown as 
function of time, where the time interval spans the main crest of the 
optical field. It therefore corresponds to the main electron burst injected 
from the tip into the junction (see Fig.~\ref{fig:Interpolated}). 
From the TDDFT, we obtain a duration of 792~as for the main burst of the electron current 
in the middle of the junction, comparable to 819~as retrieved from the nTDSE model.
}
\label{fig:Duration}
\end{figure}

At the centre of the nanogap, the polarization currents are 
exponentially small and the inter-nanowire current dominates, 
allowing its analysis. In particular, in Fig.~\ref{fig:Duration} we 
use the time evolution of the laser-induced electron current $\mathcal{I}_e(x=0,t)$ 
at the middle of the junction between the nanowires 
(see Eq.~\ref{eq:NetTransfer}) to estimate the duration of the 
most intense electron burst injected into the junction by the 
half-cycle with strongest optical field. We find that at the middle 
of the junction the electron burst propagating in the positive 
direction of the $x$-axis has a duration of 792~as (FWHM of the peak 
in the $\mathcal{I}_e(x=0,t)$ time dependence). The time delay 
between the arrival of the electron burst at the gap centre and the 
instant when the optical field reaches its extremum, stems from the 
ballistic transport of photoemitted electrons across the junction. In
Fig.~\ref{fig:Duration}, we also show the result of the nTDSE calculation.
We obtain a pulse duration of 819~as which is a bit larger than the TDDFT result. We attribute this difference primarily to the distortion of the
transport current burst by the polarisation currents in TDDFT. Obviously, field screening
and associated oscillating polarization charges are not captured by a one-electron
approach.

\begin{figure}[htb!]
\centering
\includegraphics[width=0.95\linewidth]{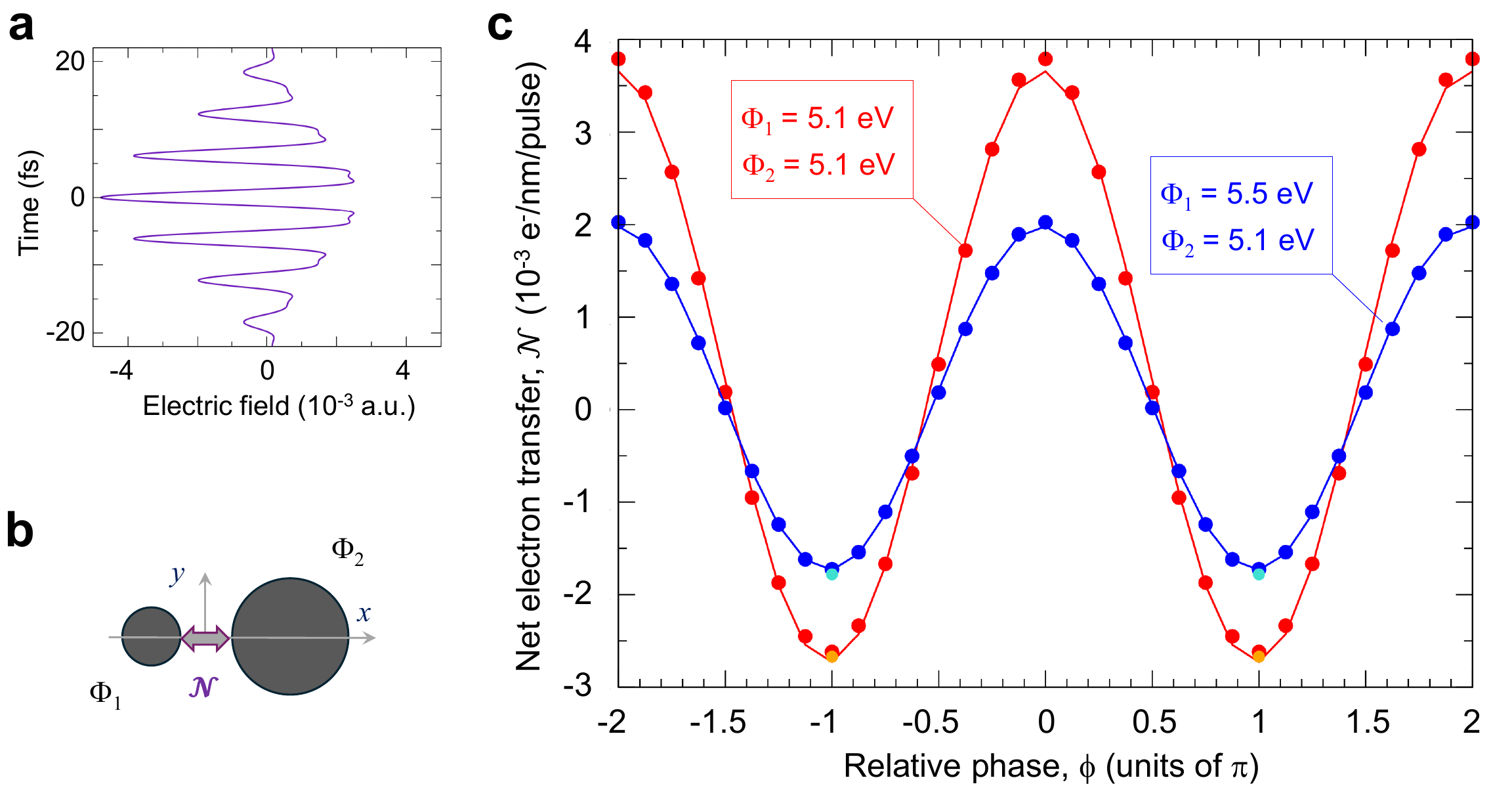}
\caption{The laser induced net electron transfer across the 1~nm gap. 
The amplitude of the incident $x$-polarized optical pulse is such that 
the maximum field reached in the gap is $10.27\,\mathrm{V\,nm}^{-1}$. 
\textbf{a}, Example of the time dependence of the electric field 
of the incidence pulse for the relative phase between 
the fundamental and second harmonic fields 
$\phi=0$ (see Eq.~\ref{Eq:ElectricField}). 
\textbf{b}, Sketch of the dimer formed by $R_1=3.3$~nm and $R_2=7$~nm  
infinite nanowires representing, respectively, the tip and the surface. 
$\Phi_{1,2}$ denotes the work function of the nanowires, and the 
double arrow depicts the electron transport.
\textbf{c}, Net electron transfer $\mathcal{N}$ defined as the number of 
electrons transferred across the gap per optical pulse and per 1~nm length 
of the dimer. TDDFT results are shown as a function of $\phi$. 
Positive $\mathcal{N}$ corresponds 
to electrons transferred from the left to the right cylinder (from the 
tip to the surface) along the positive direction of the $x$-axis. 
Dots: TDDFT calculations; lines: fit using the cosine dependence detailed 
in the main text. Red (blue) colour is used for the results obtained 
assuming $\Phi_1=\Phi_2=5.1$~eV ($\Phi_1=5.5$~eV, $\Phi_2=5.1$~eV), 
as indicated in the legend.  Orange (light blue) dots: 
results for the $\Phi_1=\Phi_2=5.1$~eV 
($\Phi_1=5.5$~eV, $\Phi_2=5.1$~eV) dimers obtained 
reversing the sign of the incident field and assuming $\phi=0$. 
If duration of the pulse is sufficient they should be equal 
to $\mathcal{N}$ values in red (blue) calculated for $\phi=\pm\pi$.
}
\label{fig:CEP}
\end{figure}

\subsection{Asymmetric two-colour delay modulation in the TDDFT model}

Finally, in Fig.~\ref{fig:CEP} we show the dependence of the net 
electron transfer $\mathcal{N}$ on the relative phase $\phi$ between the  
fundamental and second harmonic fields. For the dimer formed by cylinders 
with a work function 5.1~eV (red colour), the TDDFT results show 
a nearly ideal cosine-like dependence 
$\mathcal{N} \approx 10^{-3} \left[0.46+3.20 \cos(\phi +0.015 \pi) \right]$ 
with an asymmetry parameter $0.46/3.2=0.14$. The slight vertical offset 
stems from the symmetry breaking by the geometry of the system, with the optical near-field somewhat stronger next to the cylinder with smaller radius. It 
results in the preferential direction of electron transfer from the 
cylinder with smaller radius representing the tip to the cylinder with 
larger radius representing the surface. This is in accordance with the 3D Maxwell equations results in Section~\ref{sec:nearfield}.

In order to probe the possible effect of the difference in the work functions 
of the tip and the surface, we have also performed the calculations 
assuming $\Phi_1=5.5$~eV, $\Phi_2=5.1$~eV (results are shown with blue colour). 
In this situation the alignment of the Fermi levels of the nanowires 
is achieved by applying an external dc field. A slight overall increase 
of the tunnelling barrier leads in this situation to an overall smaller 
net electron transfer. Interestingly, since electron emission from the 
smaller cylinder ($\Phi_1=5.5$~eV) is affected stronger because of its larger  
work function, the asymmetry of the $\mathcal{N}(\phi)$ dependence is reduced. 
Here, $\mathcal{N} \approx 10^{-3} \left[0.125+1.86 \cos(\phi +0.015 \pi) \right]$ 
with an asymmetry parameter $0.125/1.86=0.07$.

The results at $\phi=\pm \pi$ shown with orange dots ($\Phi_1=\Phi_2=5.1$~eV) 
and light blue dots ($\Phi_1=5.5$~eV, $\Phi_2=5.1$~eV) are obtained by 
reversing the sign of the incident field and assuming $\phi=0$. Their 
close agreement with results obtained using field definition given 
by Eq.~\ref{Eq:ElectricField} with $\phi=\pm\pi$ affirms the convergence 
of $\mathcal{N}$ with respect to the pulse duration. 

In summary, our results show that geometric asymmetry of tip and sample 
cause asymmetry of the two-colour modulation and prefer tip-to-sample transport 
over the inverse direction if the work functions of tip and sample are similar. 
This asymmetry can be weakened (or eventually inverted) if we 
consider a higher work function of the tip with respect to the sample. 
{The tip-to-sample preferential direction of the laser-induced 
electron transport is at odds with our experimental data rather indicating 
the sample-to-tip preference. Given that an unrealistically large tip work 
function would be needed to invert the calculated tendency, we tentatively 
attribute the discrepancy to the effects not accounted for in our TDDFT 
assuming free-electron metals with ideal surfaces. 
These are (i) the local surface topography introducing sharp features and 
affecting relation between the field at the tip and at the surface; 
(ii) the difference between free-electron metal and actual band structure 
of the materials affecting reflectivity of the metal/vacuum interfaces. 
This said, while the aforementioned effects introduce certain asymmetry 
in the laser-induced current dependence on the two-colour delay, they can 
not override the symmetry breaking by the two-colour field which determines and 
controls the direction of laser-induced electron transport.
